\documentclass[]{pasj01}
\jyear{2024}
\twocolumn 
\usepackage{booktabs}
\Received{$\langle$reception date$\rangle$}
\Accepted{$\langle$acception date$\rangle$}
\Published{$\langle$publication date$\rangle$}
\usepackage{url,lscape}

\definecolor{xlinkcolor}{rgb}{0,0,1}
\definecolor{xurlcolor}{cmyk}{0.78,0.17,0.09,0} 
\makeatletter
\let\old@ssect\@ssect % Store how \@ssect
\makeatother

\usepackage{natbib,amsmath}
\usepackage[pdfpagelabels=false]{hyperref}
\hypersetup{colorlinks=True,linkcolor=red,citecolor=xlinkcolor,urlcolor=xurlcolor}

\makeatletter
\def\@ssect#1#2#3#4#5#6{%
  \NR@gettitle{#6}% Insert key \nameref title grab
  \old@ssect{#1}{#2}{#3}{#4}{#5}{#6}% Restore ifacconf's \@ssect
}
\makeatother
\usepackage[symbol]{footmisc}
\begin{document}

\title{Galaxy Morphologies Revealed with Subaru HSC and Super-Resolution Techniques II: \\ Environmental Dependence of Galaxy Mergers at $z\sim2-5$}
\author{Takatoshi \textsc{Shibuya}\altaffilmark{1}}
\author{Yohito \textsc{Ito}\altaffilmark{1}}
\author{Kenta \textsc{Asai}\altaffilmark{1, 2}}
\author{Takanobu \textsc{Kirihara}\altaffilmark{1, 3}} %[0000-0001-6503-8315]
\author{Seiji \textsc{Fujimoto}\altaffilmark{4, 5, 6, $^\dag$}} %[0000-0001-7201-5066]
\author{Yoshiki \textsc{Toba}\altaffilmark{7, 8, 9, $^\ddag$}} %[0000-0002-3531-7863]
\author{Noriaki \textsc{Miura}\altaffilmark{1}}
\author{Takuya \textsc{Umayahara}\altaffilmark{1}}
\author{Kenji \textsc{Iwadate}\altaffilmark{1}}
\author{Sadman S. \textsc{Ali}\altaffilmark{10}} %[0000-0003-3883-6500]
\author{Tadayuki \textsc{Kodama}\altaffilmark{11}}

\email{tshibuya@mail.kitami-it.ac.jp}

\altaffiltext{1}{Kitami Institute of Technology, 165, Koen-cho, Kitami, Hokkaido 090-8507, Japan}
\altaffiltext{2}{Wakayama University, 930 Sakaedani, Wakayama 640-8510, Japan}
\altaffiltext{3}{Center for Computational Sciences, University of Tsukuba, Tennodai 1-1-1, Tsukuba, Ibaraki 305-8577, Japan} 
\altaffiltext{4}{Department of Astronomy, The University of Texas at Austin, 2515 Speedway Boulevard Stop C1400, Austin, TX 78712-1205, USA}
\altaffiltext{5}{Cosmic Dawn Center (DAWN), Jagtvej 128, DK-2200 Copenhagen N, Denmark}
\altaffiltext{6}{Niels Bohr Institute, University of Copenhagen, Lyngbyvej 2, DK-2100 Copenhagen \O, Denmark}
\altaffiltext{7}{National Astronomical Observatory of Japan, 2-21-1 Osawa, Mitaka, Tokyo 181-8588, Japan}
\altaffiltext{8}{Academia Sinica Institute of Astronomy and Astrophysics, 11F of Astronomy-Mathematics Building, AS/NTU, No.1, Section 4, Roosevelt Road, Taipei 10617, Taiwan}
\altaffiltext{9}{Research Center for Space and Cosmic Evolution, Ehime University, 2-5 Bunkyo-cho, Matsuyama, Ehime 790-8577, Japan}
\altaffiltext{10}{Subaru Telescope, National Astronomical Observatory of Japan}
\altaffiltext{11}{Astronomical Institute, Tohoku University, 6-3, Aramaki, Aoba, Sendai, Miyagi 980-8578, Japan}


\maketitle

\begin{abstract}

We super-resolve the seeing-limited Subaru Hyper Suprime-Cam (HSC) images for $32,187$ galaxies at $z\sim2-5$ in three techniques, namely, the classical Richardson-Lucy (RL) point spread function (PSF) deconvolution, sparse modeling, and generative adversarial networks to investigate the environmental dependence of galaxy mergers. These three techniques generate overall similar high spatial resolution images but with some slight differences in galaxy structures, for example, more residual noises are seen in the classical RL PSF deconvolution. To alleviate disadvantages of each technique, we create combined images by averaging over the three types of super-resolution images, which result in galaxy sub-structures resembling those seen in the Hubble Space Telescope images. Using the combined super-resolution images, we measure the relative galaxy major merger fraction corrected for the chance projection effect, $f_{\rm merger}^{\rm rel,\, col}$, for galaxies in the $\sim300$ deg$^2$-area data of the HSC Strategic Survey Program and the CFHT Large Area U-band Survey. Our $f_{\rm merger}^{\rm rel,\, col}$ measurements at $z\sim3$ validate previous findings showing that $f_{\rm merger}^{\rm rel,\, col}$ is higher in regions with a higher galaxy overdensity $\delta$ at $z\sim2-3$. Thanks to the large galaxy sample, we identify a nearly linear increase in $f_{\rm merger}^{\rm rel,\, col}$ with increasing $\delta$ at $z\sim4-5$, providing the highest-$z$ observational evidence that galaxy mergers are related to $\delta$. In addition to our $f_{\rm merger}^{\rm rel,\, col}$ measurements, we find that the galaxy merger fractions in the literature also broadly align with the linear $f_{\rm merger}^{\rm rel,\, col}-\delta$ relation across a wide redshift range of $z\sim2-5$. This alignment suggests that the linear $f_{\rm merger}^{\rm rel,\, col}-\delta$ relation can serve as a valuable tool for quantitatively estimating the contributions of galaxy mergers to various environmental dependences. This super-resolution analysis can be readily applied to datasets from wide field-of-view space telescopes such as Euclid and Roman. 

\end{abstract}

\footnotetext[2]{Hubble Fellow}
\footnotetext[3]{NAOJ Fellow}

%\pagewiselinenumbers

% --------------------------------------------------
\section{Introduction}\label{sec_intro}

Galaxy evolution is thought to accelerate within dense cosmic environments such as galaxy groups and galaxy clusters \citep{2005ApJ...621..673T}. In the local Universe, old and star formation-quenched elliptical galaxies are more abundant in massive galaxy clusters than in the field environments (e.g., \citealt{1980ApJ...236..351D, 1998A&A...334...99K, 2003MNRAS.346..601G, 2004AJ....128.2677T}). This strong environmental dependence on the star formation activity and the galaxy morphology is likely to be a consequence of the accelerated galaxy evolution in galaxy overdense regions. Among proposed environmental processes to accelerate the galaxy evolution (e.g., ram-pressure stripping: \citealt{1972ApJ...176....1G}; warm and hot gas removal: \citealt{1980ApJ...237..692L, 2000ApJ...540..113B}; high speed galaxy encounters: \citealt{1996Natur.379..613M}), galaxy mergers are thought to be one of the key physical mechanisms in forming massive elliptical galaxies \citep{1994ApJ...431L...9M, 2014MNRAS.444.3357N, 2014ApJ...788L..23T}. Galaxy mergers trigger nuclear starbursts followed by the emergence and feedback of active galactic nuclei (AGN), eventually resulting in the formation of massive elliptical galaxies \citep{2008ApJS..175..356H, 2012NewAR..56...93A, 2014ApJ...782...68T}. 

The frequency of galaxy mergers in galaxy overdense regions has been extensively investigated in a large number of studies. At $z\sim0-1$, galaxy mergers and interactions occur more frequently in higher galaxy density regions due to small galaxy-galaxy separations (e.g., \citealt{2010ApJ...718.1158L, 2011arXiv1104.5470D, 2024ApJ...964L..33L}). However, the fraction of galaxy mergers in massive galaxy clusters tends to be lower than that in galaxy groups or the field environment (e.g., \citealt{2008MNRAS.388.1537M, 2008ApJ...683L..17T, 2009MNRAS.399.1157P, 2012AA...539A..46A,2023AA...679A.142O,2024arXiv240218520S}). Galaxy-galaxy interactions are suppressed due to a high velocity dispersion of $\sim500-1,000$ km/s in a deep gravitational potential well of massive galaxy clusters, as predicted in theoretical studies (e.g., \citealt{1998MNRAS.300..146G, 2001ApJ...546..223G, 2012ApJ...754...26J}). Considering the evolution of velocity dispersion, it is expected that the frequency of galaxy mergers is high in young and developing galaxy protoclusters at $z\gtrsim2$ compared to local massive galaxy clusters. 

Despite the expectation, the relation between galaxy mergers and galaxy overdensity remains controversial at $z\gtrsim2$. Using high spatial resolution data of the Hubble Space Telescope (hereafter, Hubble), some studies have indicated a high galaxy merger fraction in galaxy protoclusters (e.g., \citealt{2013ApJ...773..154L, 2016MNRAS.455.2363H, 2019ApJ...874...63W, 2023MNRAS.523.2422L}). On the other hand, there are reports suggesting that the frequency of galaxy mergers may not be correlated with galaxy overdensity \citep{2017ApJ...843..126D, 2021ApJ...919...51M, 2022arXiv221205984M}. The discrepancy might be partially due to limited galaxy samples, typically comprising $\sim100-1,000$ sources, in a study with the small field of view (FoV) of Hubble. Both wide survey area data and high spatial resolution images are needed to study the morphology of high-$z$ compact sources (i.e., merger or isolated) with statistical samples of rare galaxy overdense regions. 

To investigate galaxy merger features of rare high-$z$ objects, we have undertaken morphological studies by applying super-resolution techniques to seeing-limited images of ground-based telescopes \citep{2022PASJ...74...73S}, specifically, the Subaru/Hyper Suprime-Cam (HSC)-Strategic Survey Program (SSP) data \citep{2018PASJ...70S...4A,2018PASJ...70S...8A,2018PASJ...70S...1M,2018PASJ...70S...7C,2018PASJ...70S...3F,2018PASJ...70S...6H,2018PASJ...70...66K,2018PASJ...70S...2K,2012ApJ...756..158S,2012ApJ...750...99T,2013ApJS..205...20M}. The HSC-SSP data cover a wide area, from several hundred to $\sim1,000$ deg$^2$, providing the capability to trace diverse galaxy populations and environments, including rare high-$z$ objects of the rest-frame ultra-violet (UV) luminous galaxies and galaxy protoclusters. The pilot study (\citealt{2022PASJ...74...73S}; hereafter Paper I) has estimated the galaxy major merger fractions for galaxies at $z\sim4-7$ at the bright-end of UV luminosity functions (LFs) using the classical Richardson-Lucy (RL) point spread function (PSF) deconvolution \citep{1972JOSA...62...55R, 1974AJ.....79..745L}. The classical RL PSF deconvolution has enhanced the spatial resolution of the seeing-limited HSC images, from $\sim1^{\prime\prime}$ to $\lesssim0.\!\!^{\prime\prime}1$, which is comparable to that of Hubble. The galaxy major merger fractions estimated using the super-resolution HSC images are found to be comparable between the UV-bright galaxies and control faint sources, suggesting that galaxy major mergers are not dominant physical mechanisms for the galaxy number density excess at the bright-end of UV LFs (e.g., \citealt{2018PASJ...70S..10O, 2022ApJS..259...20H}). 

This is the second paper in a series investigating morphological properties of rare galaxy populations at high-$z$ with super-resolution techniques. In this paper, we study the environmental dependence of galaxy mergers at $z\sim2-5$ by applying several super-resolution techniques for the latest product available, the internal S21A HSC-SSP data release. Apart from the classical RL PSF deconvolution used in Paper I, we introduce two additional super-resolution techniques: sparse modeling and machine learning. While these two super-resolution techniques have been used in various domains of astronomy (e.g., \citealt{2014PASJ...66...95H, 2022PASJ...74.1329M, 2017MNRAS.467L.110S, 2021ApJ...915...71V, 2021MNRAS.504.1825S}; see also review papers of, e.g., \citealt{2023PASA...40....1H} and \citealt{2023RPPh...86g6901M}), the application to high-redshift galaxies has been relatively limited. This study has technical and scientific goals. The technical goal is to apply the three super-resolution techniques to high-$z$ galaxies and to explore the advantages and disadvantages by comparing the performance of the super-resolution. The scientific goal is to measure the galaxy merger fraction as a function of galaxy overdensity using the super-resolution HSC images. 

This study may also contribute to our understanding about physical mechanisms of environmental dependences found in high-$z$ galaxy protoclusters. Over the last two to three decades, numerous galaxy protoclusters at $z\gtrsim2$ have been discovered (e.g., \citealt{1998ApJ...492..428S, 2000A&A...361L..25P, 2002ApJ...569L..11V, 2004ApJ...605L..93S, 2005ApJ...634L.125M, 2005ApJ...620L...1O, 2006ApJ...637...58O, 2010ApJ...716.1503P, 2011Natur.470..233C, 2011MNRAS.415.2993H, 2014A&A...570A..16C, 2012ApJ...746...55T, 2012ApJ...750..137T, 2014ApJ...792...15T, 2016ApJ...826..114T, 2024MNRAS.527.6276T, 2013ApJ...767...39M, 2014A&A...572A..41L, 2018A&A...615A..77L, 2015A&A...582A..30P, 2016ApJ...823...11D, 2017ApJ...844L..23C, 2013ApJ...769...79W, 2018ApJ...856...72O, 2018ApJ...863L...3C, 2018NatAs...2..962J, 2018Natur.556..469M, 2019ApJ...879...28H, 2019ApJ...879....9S, 2020ApJ...899...79S, 2020MNRAS.496.3169A, 2022PASJ...74L..27U, 2023ApJ...945L...9I, 2023arXiv230715113F, 2023arXiv231211465S, 2023ApJ...947L..24M, 2023ApJ...955L...2H}; see also review papers of \citealt{2016A&ARv..24...14O} and \citealt{2022Univ....8..554A}, a white paper of \citealt{2019BAAS...51c.180O}, and summary tables in \citealt{2013ApJ...779..127C} and \citealt{2019ApJ...883..142H}). Within high-$z$ galaxy protoclusters, various environmental dependences have been identified. Although the galaxy mergers have been proposed as a possible physical mechanism to create the environmental dependences, the influence of galaxy mergers has not been well understood observationally. Establishing a useful relation between galaxy mergers and galaxy overdensity would allow us to discuss quantitatively the contribution of galaxy mergers to the high-$z$ environmental dependences.

This paper is structured as follows. Section \ref{sec_data} provides an overview of galaxy catalogs and images used in this study. Section \ref{sec_analysis} explains details of each super-resolution technique and methods to calculate the galaxy merger fraction and the galaxy overdensity. In Section \ref{sec_results}, we present super-resolution images and the relation between the galaxy merger fraction and the galaxy overdensity. Section \ref{sec_discussion} discusses whether the galaxy merger and overdensity relation changes or not with redshift, and effects of galaxy mergers on environmental dependences at high-$z$. Section \ref{sec_summary} summarizes our findings. 

% ---
Throughout this paper, we assume a flat universe with the cosmological parameters of $(\Omega_{\rm m}, \Omega_\lambda, h)=(0.3, 0.7, 0.7)$. All magnitudes are given in the AB system \citep{1974ApJS...27...21O, 1983ApJ...266..713O}. We refer to the Hubble F606W and F814W filters as $V_{606}$ and $I_{814}$, respectively.

% ---------- Table 
\begin{longtable}{*{6}{c}}
\caption{Limiting Magnitudes of the S21A HSC-SSP Data Used for the Super-Resolution Analysis}\label{tab_limiting_mag}
 \hline
 Layer & $g$ & $r$ & $i$ & $z$ & $y$ \\
 (1) & (2) & (3) & (4) & (5) & (6) \\
 \endhead
 \hline
  UltraDeep & $27.8$ & $27.4$ & $27.1$ & $26.8$ & $26.0$ \\ 
  Deep & $27.2$ & $26.8$ & $26.5$ & $26.0$ & $25.1$ \\ 
  Wide & $26.2$ & $25.6$ & $25.2$ & $24.7$ & $23.8$ \\  
  \hline
\multicolumn{6}{l}{(1) HSC-SSP layers. (2)-(6) Typical $5\sigma$ limiting magnitudes} \\
\multicolumn{6}{l}{measured in a $2^{\prime\prime}$-diameter circular aperture in the $g, r, i, z$, and $y$ bands. } \\
\end{longtable}
% ---------- Table 

% ---------- Table 
\begin{longtable}{*{5}{c}}
\caption{Numbers of Galaxies Used for the Super-Resolution Analysis}\label{tab_sample}
 \hline
 Layer & BX/BM$^a$ & $U$-drop$^a$ & $g$-drop$^b$ & $r$-drop$^b$ \\
         & $z\sim2$ & $z\sim3$ & $z\sim4$ & $z\sim5$ \\
 \endhead
 \hline
  UltraDeep & $4,181$ & $269$ & $29$ & $1$ \\ 
  Deep          & $15,837$ & $1,032$ & $332$ & $11$  \\ 
  Wide          & $\cdots$ & $\cdots$ & $9,748$ & $747$  \\ 
  \hline
  Total($z$) & $20,018$ & $1,301$ & $10,109$ & $759$  \\ 
  \hline
  Total & \multicolumn{4}{c}{$32,187$}  \\ 
 \hline
\multicolumn{5}{l}{$^a$ Selected from the CLAUDS photo-$z$ galaxy catalog.} \\
\multicolumn{5}{l}{$^b$ Selected from the dropout galaxy catalog.} \\
\end{longtable}
% ---------- Table 

% --------------------------------------------------
% --------------------------------------------------
% --------------------------------------------------
\section{Data}\label{sec_data}

% --------------------------------------------------
\subsection{Galaxy Catalogs}\label{sec_catalogs}

In this study, we use two catalogs of galaxies: dropout galaxies in \citet{2022ApJS..259...20H} and photo-$z$ galaxies in \citet{2023A&A...670A..82D}. These galaxy catalogs are constructed with the HSC-SSP data \citep{2018PASJ...70S...4A,2018PASJ...70S...8A,2019PASJ...71..114A}. HSC is a $\sim1.8$ deg$^2$ FoV optical camera installed on Subaru Telescope. Using the wide FoV camera of HSC, the HSC-SSP survey covers three observational layers with different survey areas and depths: the UltraDeep, Deep, and Wide layers. These layers are observed through five optical broadband filters of $g$, $r$, $i$, $z$, and $y$. Due to the data availability and the sample size for galaxy overdensity maps, we combine dropout galaxies at $z\sim4-5$ and photo-$z$ galaxies at $z\sim2-3$, as explained below.

The dropout galaxies are selected with the HSC-SSP data obtained from March 2014 through January 2018. The effective survey areas of the HSC-SSP data are $\sim3$, $\sim18$, and $\sim288$ deg$^2$ for the UltraDeep, Deep, and Wide layers, respectively, which is $\sim3$ times larger in total than that of the data used in Paper I (see also \citealt{2018PASJ...70S..10O}). Thanks to the wide survey area, a substantial number of $\sim4$ million galaxies are identified at $z\sim2-7$. Among the sample, the catalogs of $z\sim4-7$ dropout galaxies are publicly available on the Zenodo website.\footnote{https://zenodo.org/records/5512721} In this study, we use $g$-dropouts at $z\sim4$ and $r$-dropouts at $z\sim5$, the numbers of which are large enough to obtain reliable galaxy overdensity maps. The selection of the dropout galaxies is based on the Lyman break technique \citep{1996ApJ...462L..17S, 2002ARA&A..40..579G}. The color selection criteria are $g-r>1.0$, $r-i<1.0$, and $g-r>1.5(r-i)+0.8$ for $g$-dropouts, and $r-i>1.2$, $i-z<0.7$, and $r-i>1.5(i-z)+1.0$ for $r$-dropouts. See \citet{2022ApJS..259...20H} for more details. We retrieve the {\tt masked} catalogs in which sources are masked in regions affected by, e.g., halos of bright stars.

The photo-$z$ galaxies are selected in the CFHT Large Area $U$-band Deep Survey (CLAUDS;  \citealt{2019MNRAS.489.5202S}). CLAUDS is an imaging survey program with the $u$- and $u^*$-bands of an optical camera of CFHT MegaCam. The CLAUDS survey covers a $\sim20$ deg$^2$ area in the HSC-SSP UltraDeep and Deep layers. The relative survey coverage of CLAUDS is $\sim80$\% of the HSC-SSP Deep fields (see Table 2 of \citealt{2023A&A...670A..82D}). The $u$- and $u^*$-band data are useful to select $z\sim2-3$ galaxies whose Lyman break features fall in the wavelength range of $\sim3,600-4,800$\AA. Part of the CLAUDS fields is also observed in near-infrared (NIR) surveys of the VISTA Deep Extragalactic Observations (VIDEO; \citealt{2013MNRAS.428.1281J}) and UltraVISTA \citep{2012A&A...544A.156M}. Photometric redshifts are obtained through the spectral energy distribution (SED) fitting from $u$ to $y$ bands (extending to $H$ or $K$-bands for the fields with NIR data available). From the catalogs available on the CLAUDS website,\footnote{https://www.clauds.net/available-data} we select the {\tt SExtractor} catalogs with the results of the SED fitting performed using $u$ to $y$ bands for all the CLAUDS fields, which have better performance in the photo-$z$ measurements than that of the other catalogs. The precision of the photo-$z$ measurements is $\sigma\lesssim0.04$ down to the $i$-band magnitude of $i\sim25$. For further details, please refer to \citet{2023A&A...670A..82D}. 

% --------------------------------------------------
\subsection{Images for the Super-resolution Analysis}\label{sec_images}

We use the latest product available, the internal S21A HSC-SSP data release, as images for our super-resolution analysis. The S21A HSC-SSP data were obtained in observations from March in 2014 through January in 2021. The HSC images have been reduced with the {\tt hscPipe} 8.0-8.4 software (\citealt{2018PASJ...70S...5B}; see also \citealt{2010SPIE.7740E..15A, 2017ASPC..512..279J, 2019ApJ...873..111I}). The pixel scale of the HSC image is $0.\!\!^{\prime\prime}168$ per pixel. The typical seeing size of the HSC-SSP data ranges from $\sim0.\!\!^{\prime\prime}7$ to $\sim0.\!\!^{\prime\prime}8$ \citep{2022PASJ...74..247A}. Table \ref{tab_limiting_mag} presents the limiting magnitudes of the S21A HSC-SSP data. We retrieve the S21A HSC-SSP coadd cutout images by $36$ pixels $\times$ $36$ pixels ($\sim6^{\prime\prime}\times6^{\prime\prime}$) centered at the position of each galaxy. The size of the cutout images is sufficiently large to cover entire structures of $z\sim2-5$ galaxies with a typical half-light radius of $r_{\rm e}\sim0.\!\!^{\prime\prime}1-0.\!\!^{\prime\prime}2$ (e.g., \citealt{2015ApJS..219...15S}). We also download PSF images from the HSC-SSP database\footnote{https://hsc-release.mtk.nao.ac.jp/doc/index.php/tools-2/} for the analysis with the classical RL PSF deconvolution and sparse modeling. Before applying the super-resolution techniques, we change the original pixel-scale of the HSC coadd and PSF images, and define a new pixel grid to match the pixel scale of the Hubble Advanced Camera for Survey (ACS; i.e., $0.\!\!^{\prime\prime}06$ per pixel). Here we perform a linear interpolation of the flux distribution in the new pixel grid. 

% --------------------------------------------------
\subsection{Selection of Galaxies for the Super-resolution Analysis}\label{sec_selection}

Combining the two galaxy catalogs of dropout galaxies and photo-$z$ galaxies, we construct a sample of $z\gtrsim2$ galaxies for the super-resolution analysis. From the CLAUDS photo-$z$ galaxy catalog, we select galaxies whose photometric redshift falls in typical redshift ranges of $z\sim2$ BX/BM and $z\sim3$ $U$-dropout galaxies. We define galaxies with the best-fit photometric redshift {\tt Z\_BEST} of $z=1.7\pm0.3$ as BM galaxies, $z=2.2\pm0.3$ as BX galaxies, and $z=3.1\pm0.3$ as $U$-dropout galaxies. We do not apply the redshift selection for the dropout galaxies because these sources are selected in the Lyman break technique. We consider $z=3.8$ and $z=4.9$ as the representative redshifts of $g$-dropout and $r$-dropout galaxies, respectively. 

We select galaxies with the rest-frame UV continuum magnitude of $m_{\rm UV}<23.5$, which are bright enough for our super-resolution analysis. To trace the rest-frame UV continuum emission, we use magnitudes in the $g$-band for BM, the $r$-band for BX and $U$-dropout, the $i$-band for $g$-dropout, and the $z$-band for $r$-dropout galaxies. We exclude compact sources which could potentially represent AGN or stars: photo-$z$ galaxies with object classification parameters of {\tt OBJ\_TYPE}$=1$ (QSOs) or $2$ (stars), {\tt COMPACT}$=1$ (compact) in the CLAUDS photo-$z$ galaxy catalog, and dropout galaxies with the magnitude difference $m_{\rm PSF}-m_{\rm CModel}<0.15$ \citep{2019ApJ...883..183M}, where $m_{\rm PSF}$ and $m_{\rm CModel}$ are the magnitudes measured with the PSF and CModel profiles, respectively \citep{2018PASJ...70S...4A}. To further reduce contaminants, only sources flagged as {\tt MASK}$=0$ and {\tt ST\_TRAIL}$=0$ are used in the CLAUDS photo-$z$ galaxy catalog. 

Table \ref{tab_sample} presents the numbers of the selected galaxies at $z\sim2-5$. The total number of the galaxies at $z\sim2-5$ is $32,187$. Although the sample consists of two different types of galaxies, i.e., the dropout galaxies and the photo-$z$ galaxies, previous studies have found that UV LFs are in good agreement between the two types of sources (e.g., \citealt{2020MNRAS.494.1894M}). The agreement suggests that selection effects do not strongly impact the analysis of this study.

\begin{figure*}
 \begin{center}
  \includegraphics[width=150mm]{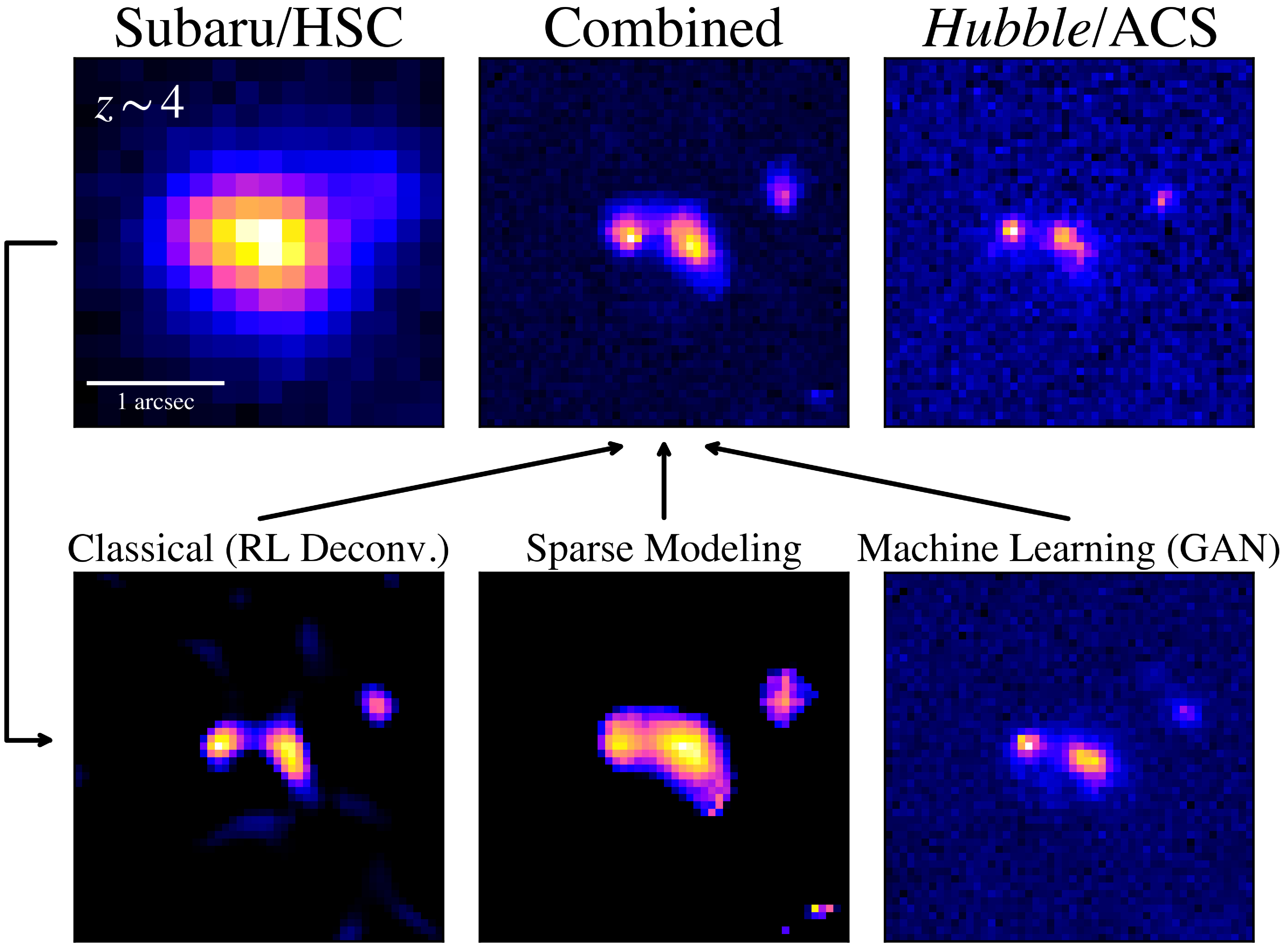}
  \end{center}
   \caption{Images of an example galaxy at $z\sim4$ analyzed in the super-resolution techniques. (Top-left) Original HSC image. (Top-middle) Combined super-resolution HSC image by averaging over the three types of super-resolution images processed in the classical RL PSF deconvolution, sparse modeling, and GAN. (Top-right) Hubble image. (Bottom) The left, middle, and right panels show the super-resolution HSC images obtained in the classical RL PSF deconvolution, sparse modeling, and GAN, respectively. The white horizontal bar indicates $1^{\prime\prime}$.}\label{fig_big_image}
\end{figure*}

\begin{figure*}
 \begin{center}
\textbf{\huge Sparse Modeling}\par\medskip
  \includegraphics[width=150mm]{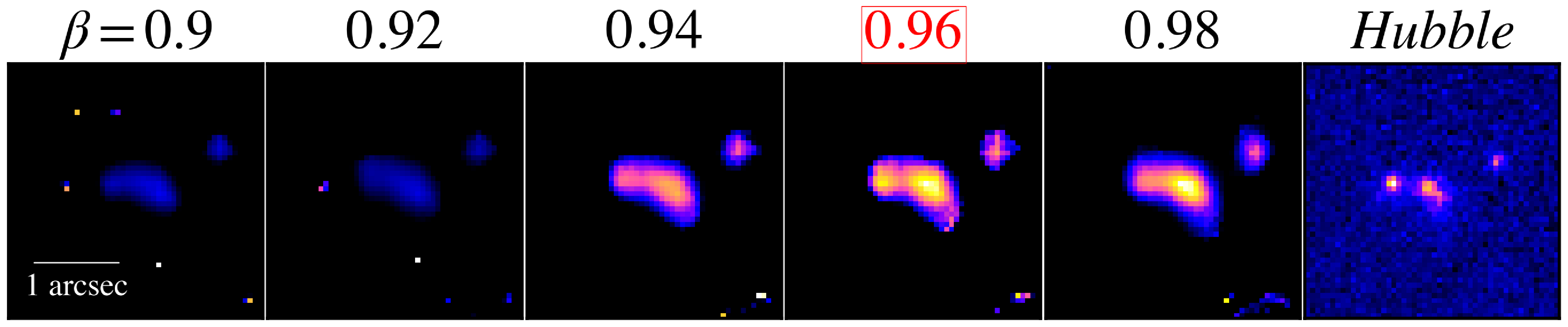}
  \mbox{}\\
  \mbox{}\\
\textbf{\huge GAN}\par\medskip
  \includegraphics[width=150mm]{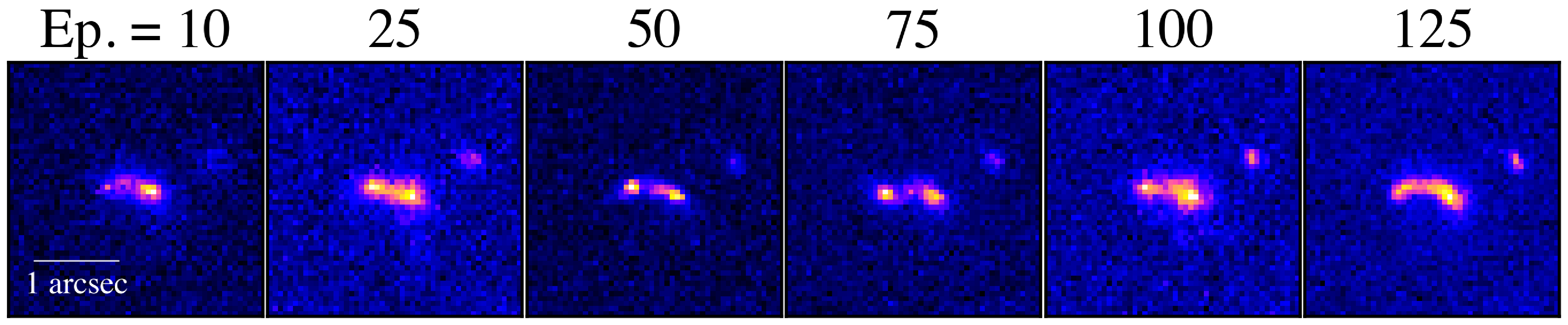}
  \includegraphics[width=150mm]{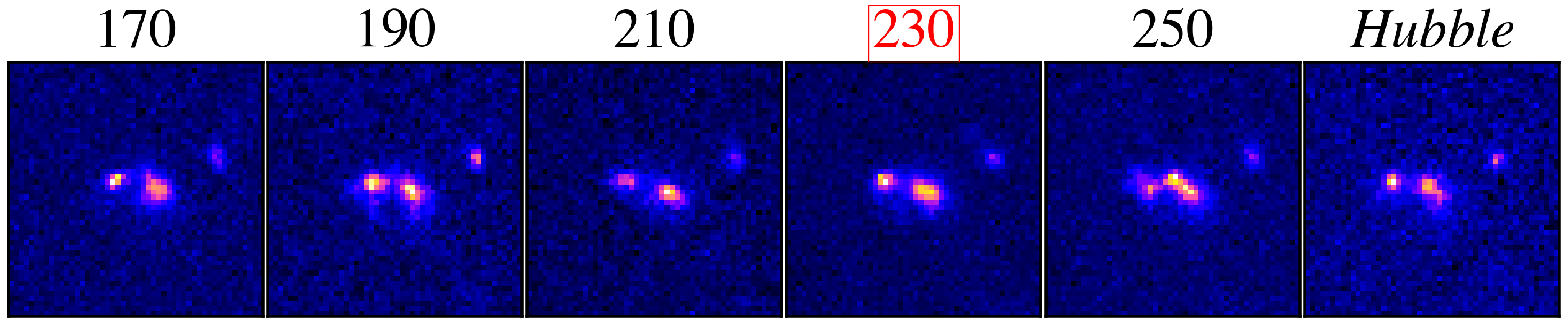}
 \end{center}
   \caption{Images of an example galaxy for determining the parameters of the sparse modeling (top panels) and GAN (bottom 2-row panels). The numbers above each image denote the hyper parameters of $\beta$ (top panels) or the epochs of GAN (bottom 2-row panels). The numbers surrounded by the red squares indicate the best hyper parameter and epoch for the super-resolution analysis. Other example galaxies are presented in Figures \ref{fig_params_sparse_others} and \ref{fig_params_gan_others} in Appendix. See Sections \ref{sec_sparse} and \ref{sec_gan} for details.}\label{fig_params_sparse_gan}
\end{figure*}

% --------------------------------------------------
\subsection{Data of Hubble Space Telescope}\label{sec_hubble}

In addition to the HSC data, we use two datasets of Hubble: COSMOS \citep{2007ApJS..172..196K} and CANDELS \citep{2011ApJS..197...35G, 2011ApJS..197...36K}. These high resolution data can serve as ground-truth images of galaxies. 

The COSMOS data is used to evaluate the performance of the super-resolution techniques. A part of the HSC-SSP UltraDeep COSMOS field has been observed with the $I_{814}$ filter of the Hubble ACS. With the Hubble COSMOS data, we create a list of galaxy mergers and isolated galaxies. Using a Hubble detection source catalog of \citet{2007ApJS..172..219L} for COSMOS and the HSC-SSP photometric redshifts obtained with the  {\tt MIZUKI} photo-$z$ software \citep{2018PASJ...70S...9T}, we select galaxy mergers and isolated galaxies at $z\gtrsim2$. We apply the same selection criteria of the flux ratio and source separation as those for the photo-$z$ and dropout galaxies, which is described in Paper I. Using the {\tt SExtractor} software \citep{1996A&AS..117..393B}, we detect sources in the super-resolution HSC images. In the source catalogs constructed with {\tt SExtractor}, galaxy major mergers are identified based on the presence of similar-flux double components. Specifically, we define galaxy close-pair systems as galaxy mergers when the systems have a double component with a flux ratio of $\mu=f_2/f_1\geq1/4$ and a source separation within $d=r_{\rm min}-r_{\rm max}=3-6.5$ kpc. Here, $f_1$ and $f_2$ are the flux values for the bright primary and the faint secondary galaxy components, respectively. The number of the selected objects is $\sim1,000$ for each class of the galaxy mergers and isolated galaxies. These objects selected in COSMOS are used to calculate the detection rates of galaxy mergers and isolated galaxies in the super-resolution HSC images (Section \ref{sec_comparison_tech}). In addition, we use these objects as the test sample to assess the generalization performance of a machine learning model in Section \ref{sec_gan}. 

On the other hand, the CANDELS data is used as the training data for a machine learning network. The detail of the training data is described in Section \ref{sec_gan}. 

% --------------------------------------------------
% --------------------------------------------------
% --------------------------------------------------
\section{Analysis}\label{sec_analysis}

% --------------------------------------------------
\subsection{Super-resolution Techniques}\label{sec_super_resolution}

We apply three super-resolution techniques: 1) the classical RL PSF deconvolution, 2) sparse modeling, and 3) machine learning. Figure \ref{fig_big_image} illustrates super-resolution HSC images of an example high-$z$ galaxy. Each technique is explained in the following three subsections (Sections \ref{sec_classical}--\ref{sec_gan}). In Section \ref{sec_comparison_tech}, we compare images generated in each super-resolution technique, and discuss the advantages and disadvantages.

% --------------------------------------------------
\subsubsection{The Classical RL PSF Deconvolution}\label{sec_classical}

The first technique is the RL PSF deconvolution \citep{1972JOSA...62...55R, 1974AJ.....79..745L}. The RL PSF deconvolution algorithm has been widely used in many applications for about five decades. Hence, we refer to the RL PSF deconvolution as the classical method in this paper. 

The classical RL PSF deconvolution is a maximum-likelihood algorithm assuming that the noise of observed images $\mathbf{y}$, 

\begin{equation}\label{eq_poisson}
\mathbf{y} = p( \mathbf{Px} ),  
\end{equation}

\noindent follows Poisson statistics $p$ at a specific image pixel $i$, 

\begin{equation}\label{eq_poisson_dist}
p(\mathbf{y}_i|\mathbf{x}) = \frac{(\mathbf{Px})_i^{\mathbf{y}_i} e^{-(\mathbf{Px})_i}}{\mathbf{y}_i!}, 
\end{equation}

\noindent where $\mathbf{P}$ is a PSF matrix, and $\mathbf{x}$ is a vector of a restored image. The classical RL PSF deconvolution has a super-resolution effect because the cutoff spatial frequency of the image $\mathbf{x}$ increases during the maximum likelihood estimation. 

We analyze the HSC images with the classical RL PSF deconvolution in the same way as that of Paper I. Specifically, we exploit the Alternating Direction Method of Multipliers (ADMM) algorithm \citep{MAL-016}. For the convergence check during the ADMM iteration, we monitor the total absolute percentage error, 

\begin{equation}\label{eq_tape}
E(n_{\rm iter}) = \sum_{i=1}^{N_{\rm im}} \frac{|P_i * x_i -y_i|}{|y_i|}, 
\end{equation}

\noindent where $*$ represents the convolution operator, $i$ denotes a specific pixel, $n_{\rm iter}$ signifies the current iteration number, and $N_{\rm im}$ means the total number of image pixels. We stop the iteration when the super-resolution image $\mathbf{x}$ satisfies a convergence criterion of $|{\rm d} E(n_{\rm iter})/ {\rm d}n_{\rm iter}| < 1\times10^{-3}$ at least $10$ times consecutively, or $n_{\rm iter}$ exceeds $N_{\rm iter}=250$. See Paper I for the detailed analysis. 

The bottom left panel of Figure \ref{fig_big_image} shows an example image processed with the classical RL PSF deconvolution technique. As demonstrated in Paper I, the classical RL PSF deconvolution reproduces substructures of high-$z$ galaxies which resemble those seen in the Hubble images. However, the classical RL PSF deconvolution tends to generate many residual noises in the restored images. In addition, the process of the RL PSF deconvolution can occasionally violate the Shannon-Nyquist sampling theorem \citep{1928PhRv...32..110N, 1949mtc..book.....S, 1998ApJ...494..472M}. To deal with these limitations, we use two additional techniques, namely, sparse modeling and machine learning.

% --------------------------------------------------
\subsubsection{Sparse Modeling}\label{sec_sparse}

The second technique is sparse modeling. The sparse modeling is an algorithm to solve problems by imposing constraints related to the sparsity in specific domains. In image restoration studies, widely utilized constraints are the sparseness in pixel values and the smoothness among adjacent pixels. These constraints can enhance the image resolution less affected by noises than the classical RL PSF deconvolution. 

In this study, we employ a method of sparse modeling proposed by \citet{2019PASJ...71...24M}. The method incorporates two additional constraints of sparseness and smoothness to the classical RL PSF deconvolution. The sparseness constraint is the Direclet function, 

\begin{equation}
  (1-\beta) \sum_u\log\rho(u), \hspace{1em} \beta>0, 
\end{equation}

\noindent where $\rho$ is an image normalized with the total flux, $\rho=y(u)/\Sigma_u{y(u)}$, $u$ is a pixel index, and $\beta$ is a hyper parameter controlling the degree of sparseness. A smaller $\beta$ value generates a more sparse image. 

The smoothness constraint is the total squared variation (TSV), 

\begin{eqnarray}
  V({\rho}) & = & \sum_{i = 1}^{m - 1} \sum_{j = 1}^{n - 1}
  \left[ (\rho_{i,j} - \rho_{i + 1,j})^2
    + (\rho_{i,j} - \rho_{i,j + 1})^2 
    \right] \\
  & & + \sum_{i = 1}^{m - 1} (\rho_{i,n} - \rho_{i + 1,n})^2
  + \sum_{j = 1}^{n - 1} (\rho_{m,j} - \rho_{m, j+1})^2, 
\end{eqnarray}

\noindent where $\mu$ is a hyper parameter controlling the degree of smoothness. A higher $\mu$ value generates a smoother image. The TSV is a squared version of the total variation (TV; \citealt{1992PhyD...60..259R}) and has been introduced to resolve radio interferometer images taken with Event Horizon Telescope (EHT; \citealt{2018ApJ...858...56K}) and ALMA \citep{2020ApJ...895...84Y}. While TV creates mosaic-like structures, TSV restores more realistic images for smooth objects.

% ---------- Table 
\begin{longtable}{*{8}{c}}
\caption{Magnitude Limits and Training Data for GAN}\label{tab_training_data}
 \hline
  Layer & \multicolumn{2}{c}{COSMOS} & \multicolumn{2}{c}{UDS} & \multicolumn{2}{c}{AEGIS} & $N_{\rm train}$ \\
 \cmidrule(lr){2-3} \cmidrule(lr){4-5} \cmidrule(lr){6-7}
         & $V_{606}$ & $I_{814}$ & $V_{606}$ & $I_{814}$ & $V_{606}$ & $I_{814}$ &  \\
 (1) & (2) & (3) & (4) & (5) & (6) & (7) & (8) \\
 \endhead
 \hline
  UltraDeep & $25.5$ & $25.3$ & $25.6$ & $25.6$ & $\cdots$ & $\cdots$ & $35,228$ \\ 
  Deep & $25.0$ & $24.7$ & $25.1$ & $25.1$ & $\cdots$ & $\cdots$ & $23,484$ \\ 
  Wide & $24.5$ & $24.3$ & $24.6$ & $24.6$ & $24.6$ & $24.2$ & $25,365$ \\ 
  \hline
\multicolumn{8}{l}{(1) HSC-SSP Layers. } \\
\multicolumn{8}{l}{(2), (3) Magnitude limits to select galaxies in COSMOS in the $V_{606}$ and $I_{814}$ bands, respectively. } \\
\multicolumn{8}{l}{(4), (5) Magnitude limits to select galaxies in UDS in the $V_{606}$ and $I_{814}$ bands, respectively. } \\
\multicolumn{8}{l}{(6), (7) Magnitude limits to select galaxies in AEGIS in the $V_{606}$ and $I_{814}$ bands, respectively. } \\
\multicolumn{8}{l}{(8) Total numbers of galaxies in the training sample for GAN. } \\
\end{longtable}
% ---------- Table 

By adding the sparseness and smoothness constraints to the classical RL PSF deconvolution, the cost function is defined as, 

\begin{equation}
  L(\rho) = - L_\rho(\rho) + (1 - \beta) \sum_u \log \rho(u) + \mu V(\rho),
\end{equation}

\noindent where $L_\rho(\rho) = \sum_{v} Y(v) \log \left[ \sum_u t(v, u) \rho(u) \right]$. The joint distribution of $Y(v)$ is a product of two functions: the Poisson and multinomial functions. The optimization problem aims to minimize the cost function with the hyper parameters of $\beta$ and $\mu$. See \citet{2019PASJ...71...24M} and \citet{2014NIMPA.760...46I} for details of the optimization process. 

The bottom middle panel of Figure \ref{fig_big_image} displays the image analyzed with the sparse modeling technique. We determine the hyper parameters of sparseness $\beta$ and smoothness $\mu$ through a grid search of $\beta$ and $\mu$ by visually inspecting restored images for several randomly-chosen sources. The top panel of Figure \ref{fig_params_sparse_gan} shows images at different hyper parameters of $\beta$. Other example galaxies are presented in Figure \ref{fig_params_sparse_others} in Appendix. Here we only present the dependence on $\beta$ because we find that $\mu=0$ offers the most optimal results in a grid search of the hyper parameters. The parameter of $\beta=0.96$ produces an image which most closely resembles that of Hubble. Hence, we adopt $\beta=0.96$ and $\mu=0$ as the hyper parameters for the super-resolution analysis. 

\begin{figure*}
 \begin{center}
 \includegraphics[width=170mm]{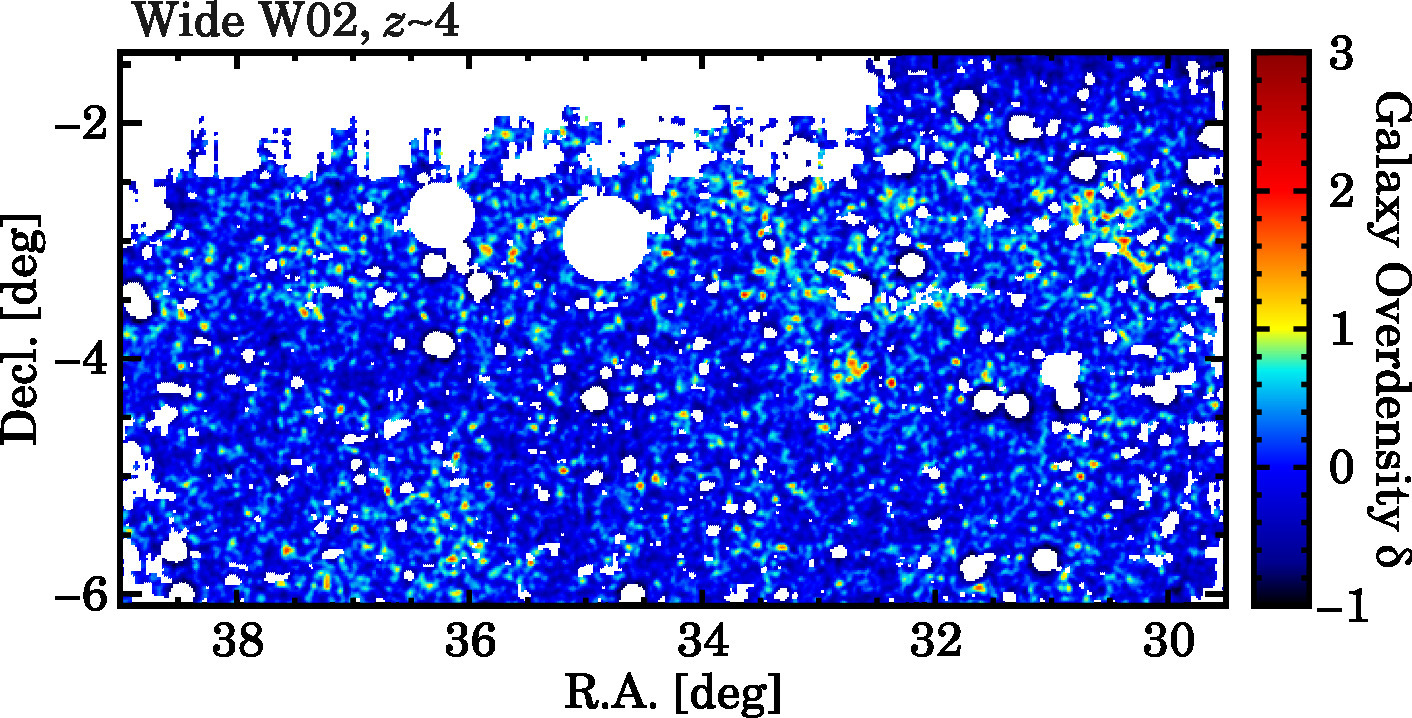}
 \end{center}
   \caption{Example of the galaxy overdensity maps, $z\sim4$ galaxies in the Wide W02 field. The higher density $\delta$ regions are indicated by redder colors. The white areas depict regions that are masked due to survey edges, bright stars, or insufficient survey depth.}\label{fig_overdensity}
\end{figure*}

% --------------------------------------------------
\subsubsection{Generative Adversarial Network}\label{sec_gan}

The third technique is based on machine learning, specifically, generative adversarial network (GAN; \citealt{2014arXiv1406.2661G}). The GAN is a machine learning algorithm designed for data generation tasks. The GAN consists of two networks: the generator and the discriminator. The generator yields fake images by learning the characteristics of numerous input real images, while the discriminator distinguishes real images from generated (fake) images. After the successful training, the GAN is able to generate synthetic images which are closely similar to real images. 

In this study, we employ the conditional GAN (cGAN; \citealt{2014arXiv1411.1784M}). The cGAN framework trains the generator and discriminator incorporating additional conditions, such as class labels and descriptive tags. For our task of generating high-resolution images from low-resolution inputs, we use a specific cGAN known as pix2pix, which conditions on images \citep{2016arXiv161107004I, isola2017image}. By learning a mapping from input images to output images, the pix2pix cGAN performs image-to-image translation tasks. Through training with a large dataset of image pairs consisting of low-resolution and high-resolution images, the network can effectively convert low-resolution inputs to high-resolution outputs. For our super-resolution analysis, we adopt the HSC (Hubble) images as low-resolution (high-resolution) images. 

To build our training samples, we select galaxies in the HSC-SSP fields of UltraDeep COSMOS, UDS, and Wide AEGIS, which are partially observed in the Hubble CANDELS survey. Galaxies for the training sample are extracted from the 3D-HST galaxy catalog \citep{2014ApJS..214...24S}. We use galaxies brighter than the $15\sigma$ limiting magnitudes of the CANDELS fields. Table \ref{tab_training_data} summarizes the magnitude limits used for the galaxy selection in each HSC-SSP layer. Due to the lack of Deep-depth HSC-SSP images in the Hubble CANDELS fields, we substitute images from the previous S18A UltraDeep release, exhibiting similar depths to those of the S21A Deep layer. To augment the training sample, we utilize images from two Hubble ACS bands, $V_{606}$ and $I_{814}$, as high-resolution images. Correspondingly, we employ the HSC $r$ and $i$ band images, covering wavelengths similar to those of $V_{606}$ and $I_{814}$, respectively, as low-resolution images. These images are paired as $r$--$V_{606}$ and $i$--$I_{814}$. To avoid redundantly inputting images with similar galaxy structures, we rotate the $r$ and $V_{606}$ images by $90$ degrees from the original orientation. Table \ref{tab_training_data} presents the number of galaxies in the training sample. The training sample does not include the sources in the test sample (Section \ref{sec_hubble}), which is used to assess the generalization performance of GAN. The total numbers of galaxies in the training sample are $N_{\rm train}\sim20,000-35,000$, which are roughly one order of magnitude larger than dataset in previous studies ($N_{\rm train}\sim600-4,000$; e.g., \citealt{2017MNRAS.467L.110S, 2021arXiv210309711G}). Although the number of the galaxies for GAN may seem large,  these galaxies are identified in small areas of the CANDELS fields, covering only $\sim0.16$ deg$^2$ \citep{2014ApJS..214...24S}, which is significantly smaller than the $\sim300$ deg$^2$ of the HSC-SSP survey. By combining the HSC-SSP survey data and the super-resolution, we can study the relation between the galaxy mergers and the galaxy overdensity with large galaxy samples.

We use a publicly available {\tt PyTorch} code of pix2pix cGAN.\footnote{https://github.com/junyanz/pytorch-CycleGAN-and-pix2pix} We set the default architectures of {\tt PatchGAN} and {\tt resnet\_9blocks} as the discriminator and the generator, respectively. During the training, we augment the input data by flipping images randomly along the horizontal axis, which doubles the size of the training sample. The training time takes $\sim3-5$ days for each HSC-SSP layer using GPU NVIDIA Quadro P5000. 

To determine the optimal epochs for generating high-resolution images, we visually inspect images at various epochs for several randomly-chosen sources during the training process. The bottom panels of Figure \ref{fig_params_sparse_gan} present images at different epochs of the GAN training process. Other example galaxies are presented in Figure \ref{fig_params_gan_others} in Appendix. We find that the most favorable epochs are $230$ for UltraDeep, $250$ for Deep, and $270$ for Wide. As shown in Figure \ref{fig_params_sparse_gan}, similarly good images are output at different epochs (e.g., ${\rm Epochs}=170$ and $210$).  However, note that the scientific goal of this study is to measure the relative galaxy merger fraction in galaxy overdense regions and the field environment. Even if we fail to determine the best hyper parameters and ideal training epochs, we are able to achieve this scientific goal. By conducting the super-resolution analysis with consistent hyper parameters and training epochs, we can compare the galaxy merger fractions at different levels of galaxy overdensity under the same conditions. 

The bottom right panel of Figure \ref{fig_big_image} shows the example image generated with GAN.

% --------------------------------------------------
\subsection{Calculation of Galaxy Overdensity}\label{sec_overdensity}

To investigate the environmental dependence of galaxy mergers, we calculate the galaxy overdensity $\delta$ in the same manner as those of \citet{2016ApJ...826..114T} and \citet{2018PASJ...70S..12T}. The galaxy overdensity is defined as the excess in the local surface number density of galaxies, 

\begin{equation}
 \delta = \frac{N_{\rm gal} - \overline{N_{\rm gal}}}{\overline{N_{\rm gal}}}
\end{equation}

\noindent where $N_{\rm gal}$ and $\overline{N_{\rm gal}}$ are the number of galaxies within a galaxy search aperture and its average, respectively. The radius of the galaxy search aperture is set to 0.75 physical Mpc, which corresponds to $1.\!\!^{\prime}5$, $1.\!\!^{\prime}6$, $1.\!\!^{\prime}8$, and $1.\!\!^{\prime}9$ for galaxies at $z\sim2$, $z\sim3$, $z\sim4$, and $z\sim5$, respectively. The galaxy search apertures are distributed in a grid pattern with $1^\prime$ intervals. At each grid, we calculate $\delta$ with the photo-$z$ and dropout galaxies with $m_{\rm UV}\lesssim25-26$. To estimate the area of sky regions affected by halos of bright stars in the galaxy search apertures, we use the random catalogs provided on the HSC-SSP database. The random catalogs contain randomly distributed data points over the HSC-SSP fields at a specific number density, which enables us to estimate effective survey areas and to identify problematic regions \citep{2019PASJ...71..114A}. We use only galaxy search apertures whose masked area is narrower than 5\% for the calculation of $\delta$. Apertures with a $>50$\%-masked area are excluded from the analysis of the galaxy mergers and galaxy overdensity. After drawing the galaxy overdensity maps, we manually remove peculiar overdense regions which are likely to be caused by, e.g., the unmasked light near bright stars. Figure \ref{fig_overdensity} shows an example of the galaxy overdensity maps.

\begin{figure*}
 \begin{center}
  \includegraphics[width=170mm]{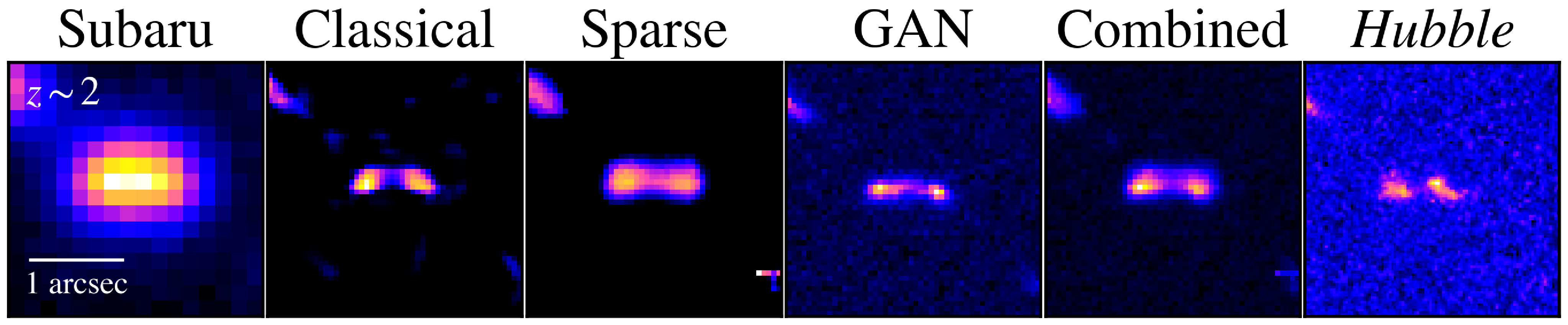}\\
  \includegraphics[width=170mm]{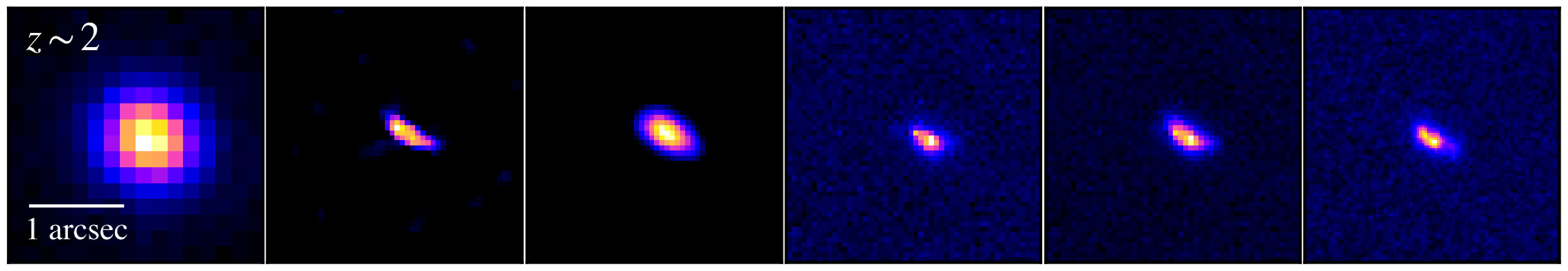}\\
  \includegraphics[width=170mm]{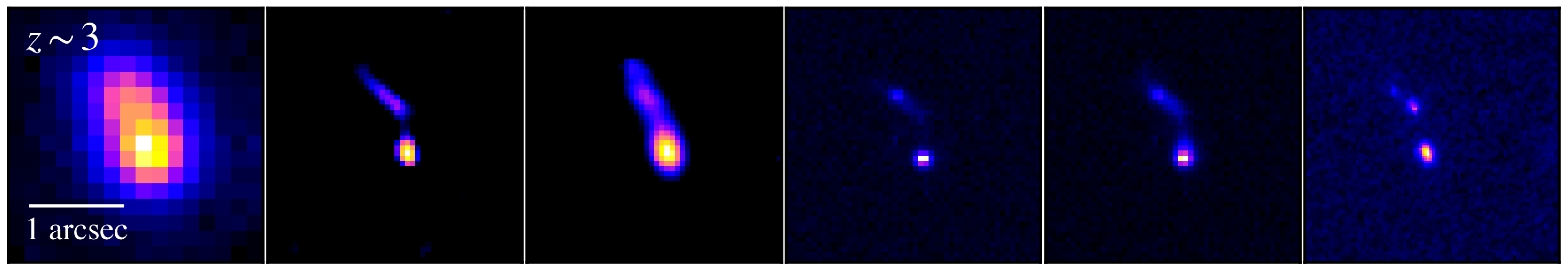}\\
  \includegraphics[width=170mm]{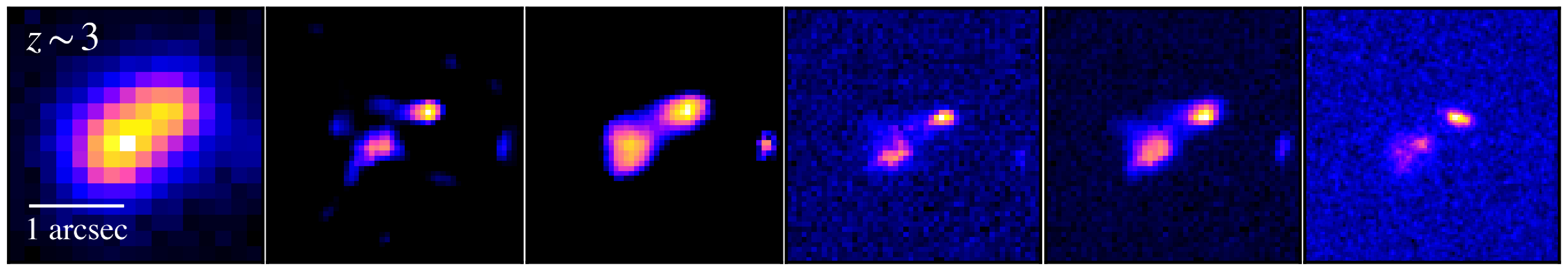}\\
  \includegraphics[width=170mm]{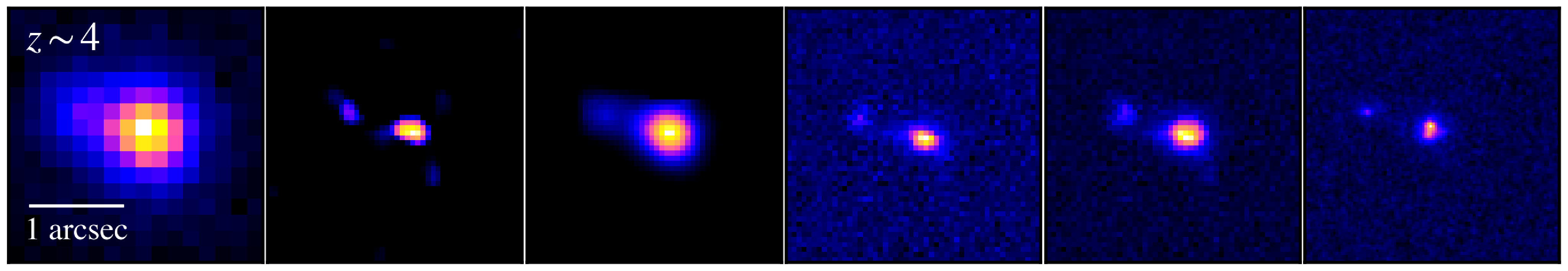}\\
  \includegraphics[width=170mm]{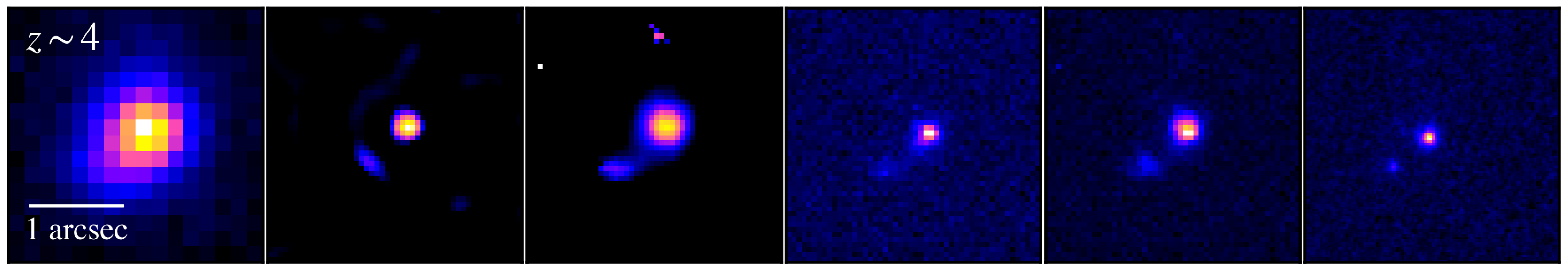}\\
  \includegraphics[width=170mm]{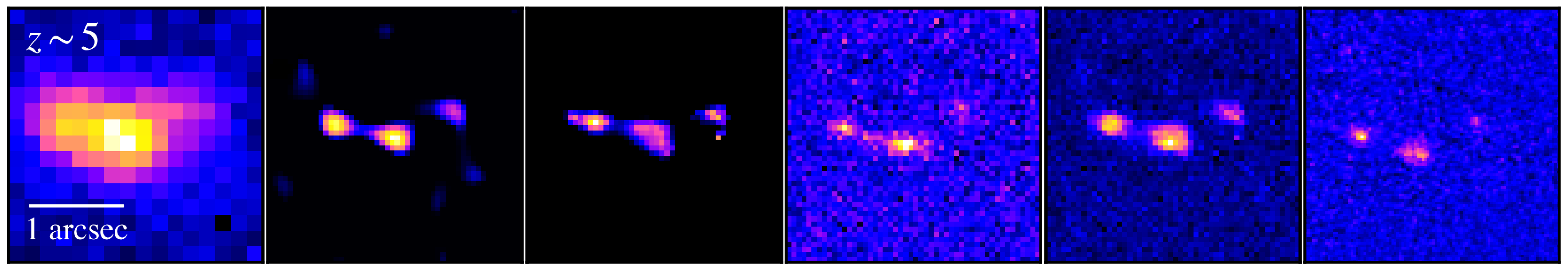}\\
  \end{center}
   \caption{Same as Figure \ref{fig_big_image}, but for other example galaxies at $z\sim2-5$. From left to right, the original HSC images, the three super-resolution HSC images processed in the classical RL PSF deconvolution, sparse modeling, and GAN, the super-resolution HSC images by combining the three super-resolution images, and the Hubble images. The redshift of each galaxy is shown at the top left of each panel.}\label{fig_other_imgs_hubble}
\end{figure*}

\begin{figure*}[t!]
 \begin{center}
  \includegraphics[height=80mm]{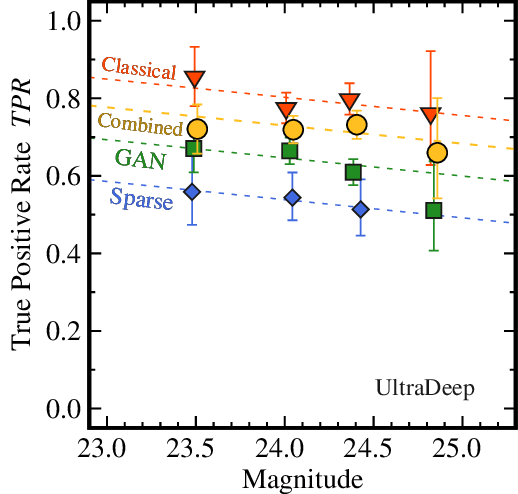}
  \mbox{}
  \includegraphics[height=80mm]{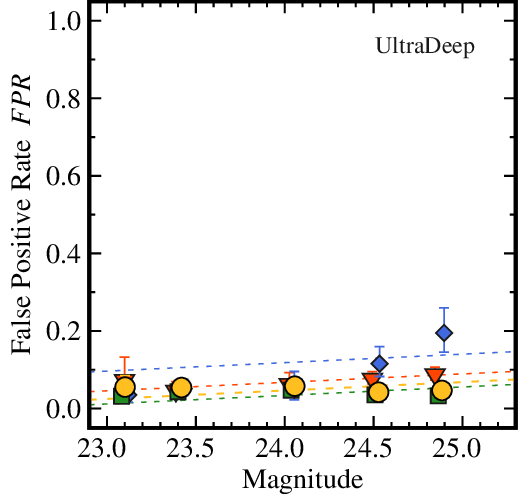}
  \mbox{}\\
  \mbox{}\\
  \includegraphics[height=80mm]{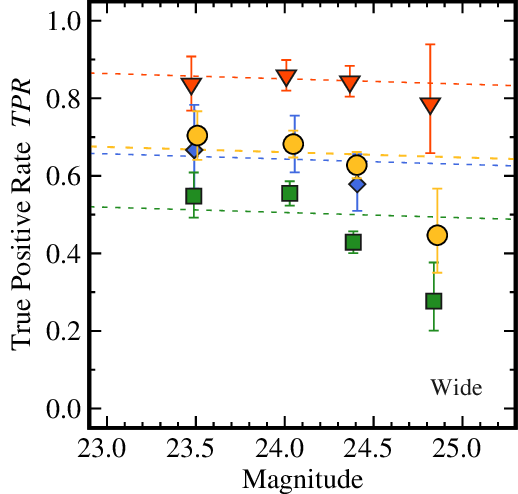}
  \mbox{}
  \includegraphics[height=80mm]{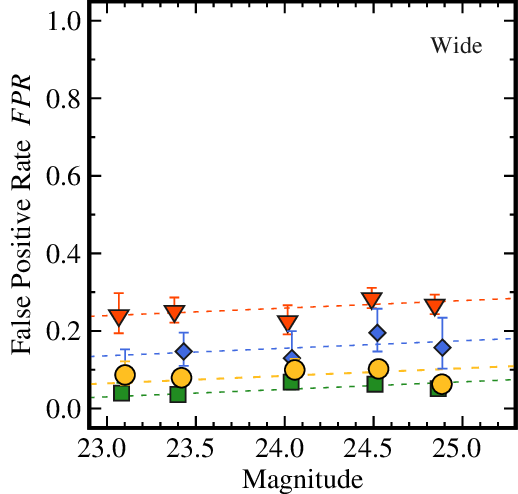}
  \end{center}
   \caption{True positive rate ($\mathit{TPR}$, left) and false positive rate ($\mathit{FPR}$, right) as a function of magnitude. The top and bottom panels denote the UltraDeep and Wide fields, respectively. The symbols indicate $\mathit{TPR}$ and $\mathit{FPR}$ estimated using the images of the classical RL PSF deconvolution (red inverse triangles), sparse modeling (blue diamonds), GAN (green squares), and the combined images (yellow circles). The panels show $\mathit{TPR}$ and $\mathit{FPR}$ estimated with at least $10$ galaxies in each magnitude bin. The error bars of each symbol are calculated based on Poisson statistics from the galaxy number counts. The red dashed lines depict the best-fit linear functions to the measurements with the classical RL PSF deconvolution. With the slope fixed, the best-fit linear functions are shifted to represent the measurements with the other super-resolution techniques (blue: sparse; green: GAN; yellow: combined).}\label{fig_tpr_fpr}
\end{figure*}

% --------------------------------------------------
% --------------------------------------------------
% --------------------------------------------------
\section{Results}\label{sec_results}

% --------------------------------------------------
\subsection{Comparison of the Super-resolution Techniques}\label{sec_comparison_tech}

We compare the performance of the super-resolution techniques. Figure \ref{fig_other_imgs_hubble} shows high-resolution images obtained in the three super-resolution techniques for galaxies at $z\sim2-5$. Basically, all the super-resolution techniques reproduce the galaxy shapes which resemble those seen in the Hubble images. The super-resolution techniques clearly reveal sub-structures of galaxies, e.g., individual components within galaxy mergers, at a scale smaller than the full width at half maximum (FWHM) of the HSC PSF (i.e., $\sim0.\!\!^{\prime\prime}6-1.\!\!^{\prime\prime}0$). Although the three techniques generate overall similar high spatial resolution images, there are advantages and disadvantages in each super-resolution technique. The classical RL PSF deconvolution excels in resolving galaxy sub-structures with close separations of $\sim0.\!\!^{\prime\prime}1-0.\!\!^{\prime\prime}2$. The restored images indicate that the classical RL PSF deconvolution has the highest resolving power in the super-resolution techniques. However, many residual noises are seen in the images of the classical RL PSF deconvolution. The sparse modeling significantly reduces the noises thanks to the sparseness and smoothness constraints, while  the sparse modeling tends to resolve galaxy close-pairs less clearly than the classical RL PSF deconvolution. The GAN machine learning technique generates realistic images with the sky background noises similar to those in the Hubble images. Another advantage of GAN is a short amount of computing time for generating images, $\lesssim0.01$ seconds per image. Despite these merits, GAN requires the large dataset of HSC-Hubble image pairs to efficiently train the network.  

To check the performance of each high-resolution technique quantitatively, we calculate the true positive rate ($\mathit{TPR}$) and the false positive rate ($\mathit{FPR}$) in identifying galaxy mergers. The $\mathit{TPR}$ is defined as 

\begin{equation}\label{eq_tpr}
  TPR = \frac{N_{\rm true}}{N_{\rm real\mathchar`-merger}}
\end{equation}

\noindent where $N_{\rm true}$ is the number of objects that are correctly selected as galaxy mergers in our super-resolution analysis (i.e., the classical, sparse, GAN, or combined), and $N_{\rm real\mathchar`-merger}$ is the total number of real galaxy mergers selected with the Hubble images. The $\mathit{FPR}$ is defined as 

\begin{equation}\label{eq_fpr}
  FPR = \frac{N_{\rm false}}{N_{\rm isolated}}
\end{equation}

\noindent where $N_{\rm false}$ is the number of isolated galaxies that are incorrectly selected as galaxy mergers in our super-resolution analysis, and $N_{\rm isolated}$ is the total number of isolated galaxies selected with the Hubble images. To obtain $N_{\rm true}$ and $N_{\rm false}$, we select galaxy mergers and isolated galaxies with the super-resolution images in the same manners as those in Section \ref{sec_images}. 

Figure \ref{fig_tpr_fpr} shows $\mathit{TPR}$ and $\mathit{FPR}$ as a function of magnitude. For all the super-resolution techniques, we find that the $\mathit{TPR}$ ($\mathit{FPR}$) tends to increase (decrease) gradually toward bright magnitudes. The $\mathit{TPR}$ ($\mathit{FPR}$) values in the UltraDeep fields are typically higher (lower) than those in the Wide fields. In all of the super-resolution techniques, the classical RL PSF deconvolution has the highest $\mathit{TPR}$ and $\mathit{FPR}$ values, which are consistent with the highest resolving power and the most noisy images, respectively. On the other hand, the sparse modeling and GAN suppress $\mathit{FPR}$ compared to that of the classical RL PSF deconvolution, especially in the Wide fields. However, the sparse modeling and GAN show low $\mathit{TPR}$ values. These $\mathit{TPR}$ and $\mathit{FPR}$ measurements confirm the qualitative evaluation of image characteristics in each super-resolution technique.

To obtain average features of restored galaxy images and to alleviate disadvantages associated with the individual techniques, we generate a combined image by averaging over the three types of super-resolution images. The top central panel of Figure \ref{fig_big_image} shows an example of the combined images. Such an approach has been applied to data of the black hole shadows taken with EHT \citep{2019ApJ...875L...1E, 2019ApJ...875L...4E}. Before combining the images, we normalize pixel values of each image to match the variations in the flux scale among the super-resolution images. To achieve this, we measure the total flux around the central regions of $25$ pixels $\times$ $25$ pixels ($4.\!\!^{\prime\prime}2\times4.\!\!^{\prime\prime}2$), which sufficiently cover main components of galaxies. In the measurements, we exclude pixels with a pixel value with zero- and negative-flux because images obtained in the sparse modeling technique have many zero-value pixels. By summing the three types of the pixel value-normalized super-resolution images, we obtain the average images. 

As shown in Figure \ref{fig_tpr_fpr}, the combined images have the ability to identify galaxy mergers with a relatively high $\mathit{TPR}$ of $\gtrsim70-80$\% and a low $\mathit{FPR}$ of $\lesssim5-10$\% at bright magnitudes $m\lesssim23.5$. As our sample galaxies are bright, i.e., $m<23.5$, the galaxy merger fraction can be estimated at high $\mathit{TPR}$ and low $\mathit{FPR}$ values. We adopt the combined images as the representative super-resolution images for the galaxy merger fractions. 

To make the analysis simple and to reduce potential systematic errors, we do not correct for $\mathit{TPR}$ and $\mathit{FPR}$ to calculate the galaxy merger fractions in this study. In Paper I, we correct for these TPR and FPR values in order to measure the absolute values of the galaxy merger fraction. The correction must be applied because the main goal of Paper I is to compare the merger fractions between bright and faint galaxies. On the other hand, the primary focus of this paper is to examine the relative trends of the galaxy merger fraction in different environments. Measuring the galaxy merger fractions for galaxies at a fixed redshift and comparable magnitudes ensures a consistent evaluation of the environmental dependence on galaxy mergers. The effect of magnitude difference on the galaxy merger fraction is discussed in Section \ref{sec_real}. 

\begin{figure*}
 \begin{center}
  \includegraphics[width=150mm]{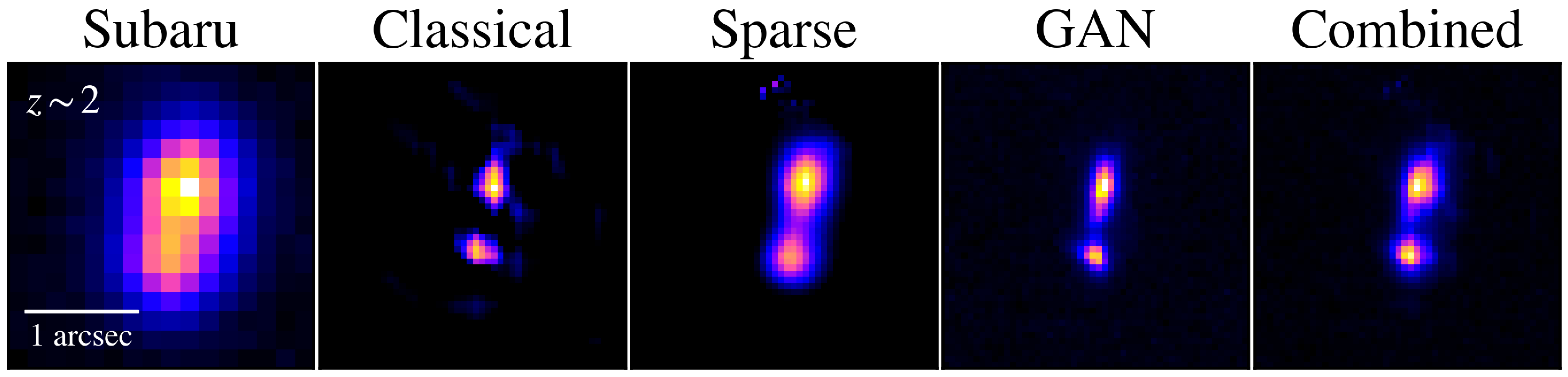}
  \includegraphics[width=150mm]{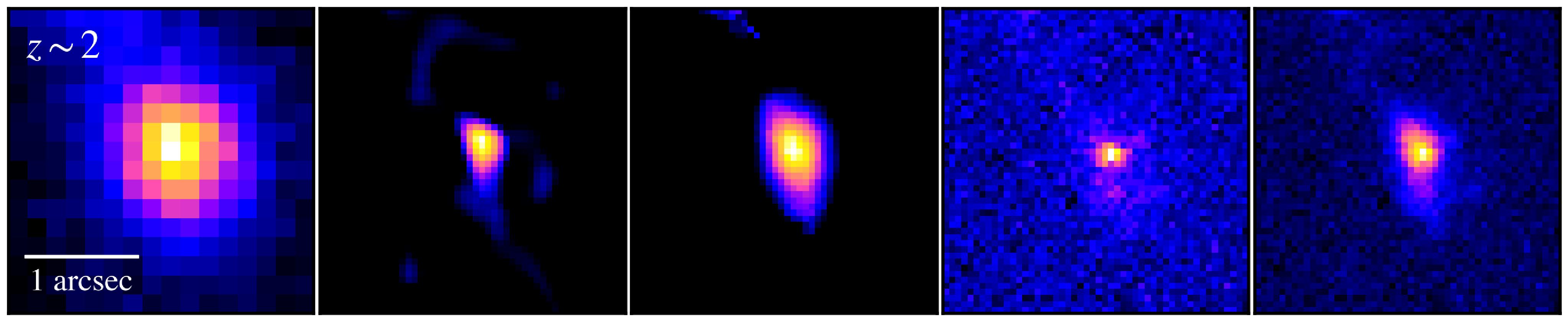}
  \includegraphics[width=150mm]{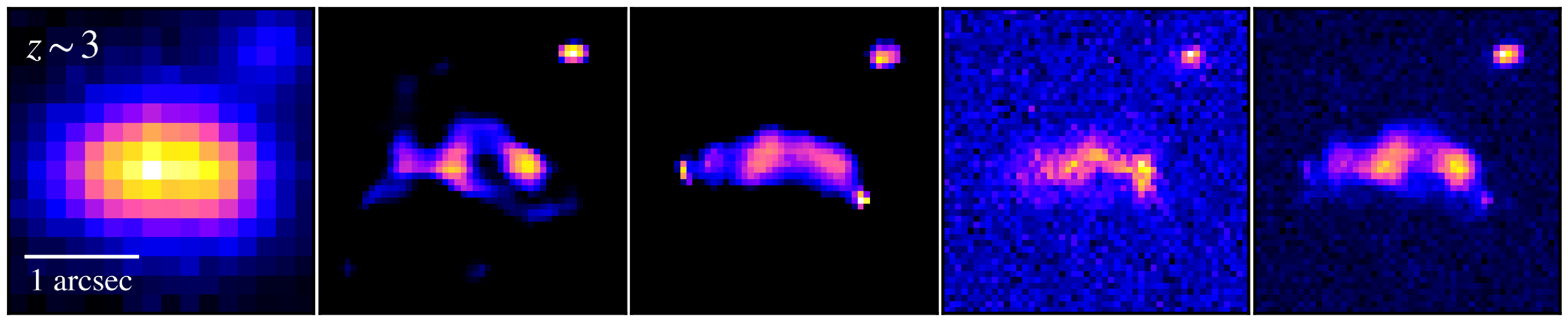}
  \includegraphics[width=150mm]{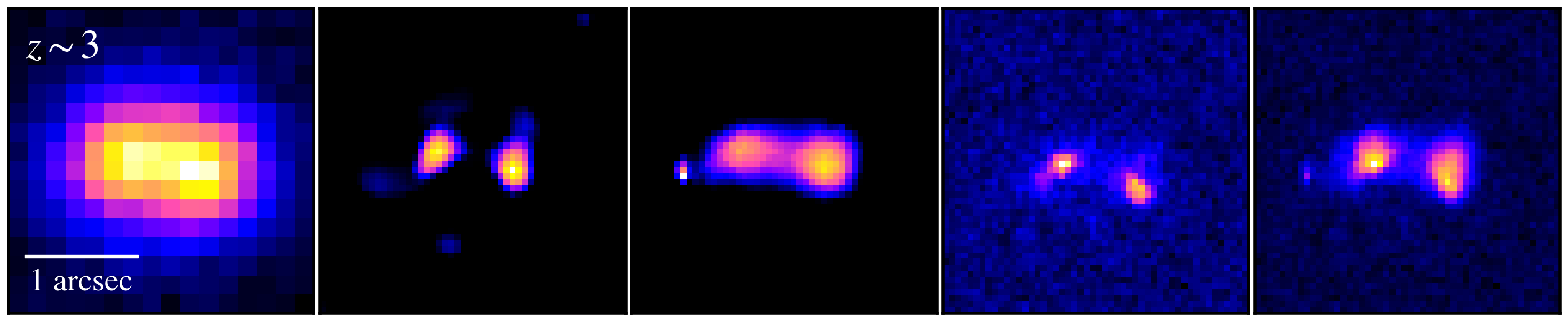}
  \includegraphics[width=150mm]{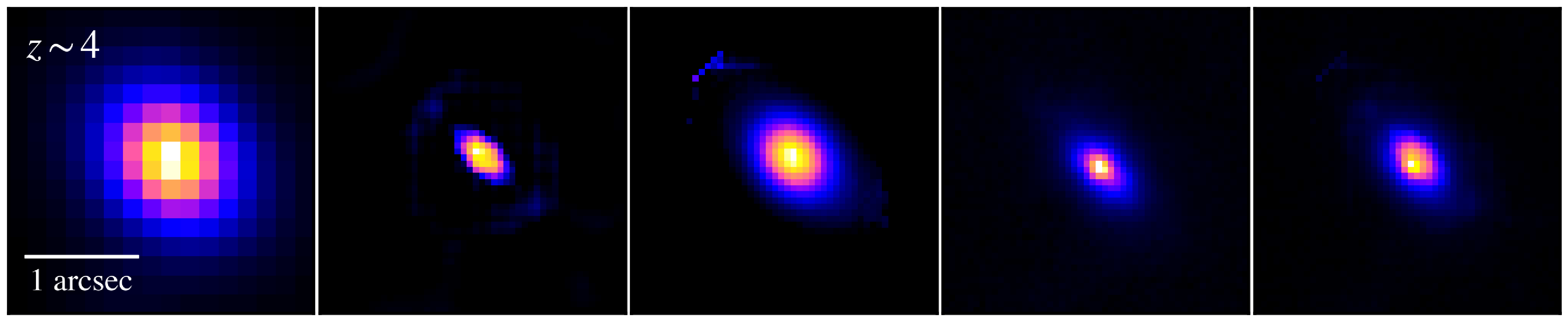}
  \includegraphics[width=150mm]{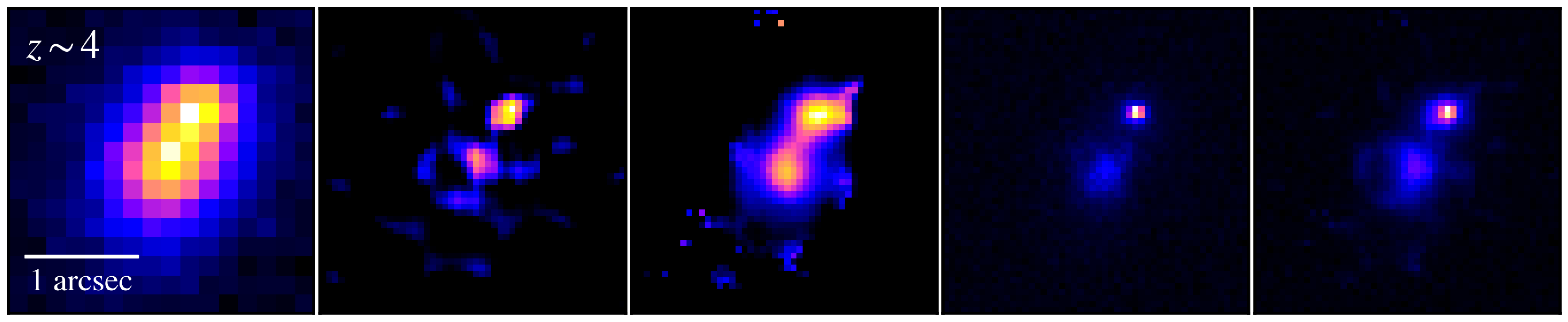}
  \includegraphics[width=150mm]{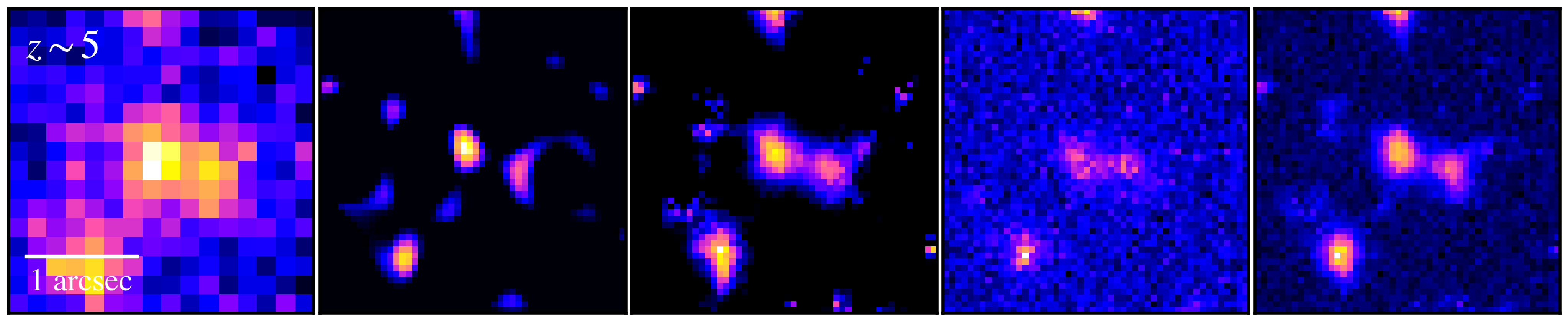}
 \end{center}
   \caption{Images of example galaxies at $z\sim2-5$ analyzed in our super-resolution technique. From left to right, the original HSC images, the three super-resolution HSC images obtained in the classical RL PSF deconvolution, sparse modeling, GAN, the super-resolution HSC images by combining the three super-resolution images. The redshift of each galaxy is shown at the top left of each panel. The white horizontal bar indicates $1^{\prime\prime}$.}\label{fig_imgs}
\end{figure*}

\begin{figure*}
 \begin{center}
  \includegraphics[width=150mm]{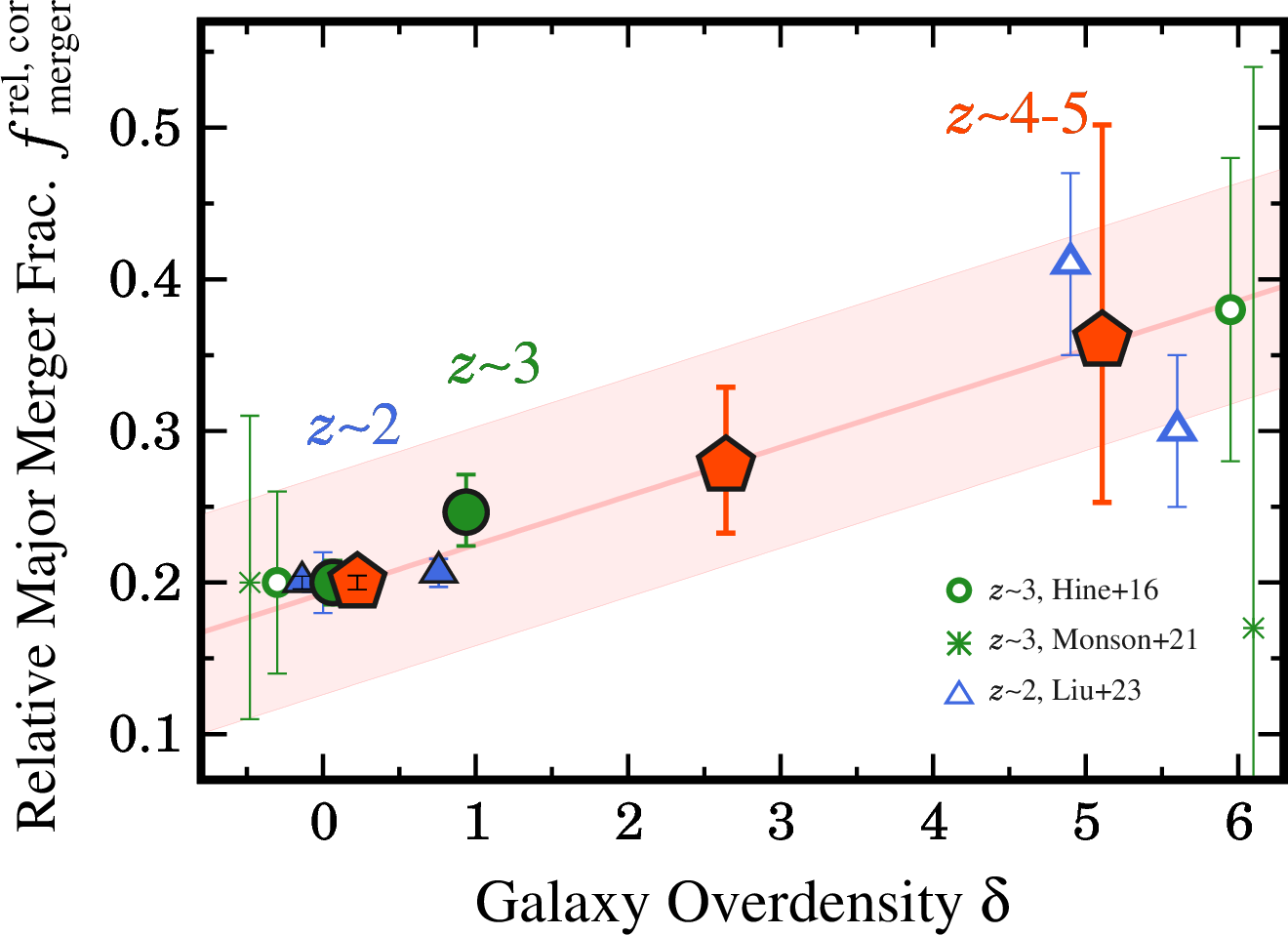}
 \end{center}
   \caption{Relative galaxy merger fraction $f_{\rm merger}^{\rm rel,\, cor}$ as a function of galaxy overdensity $\delta$ at $z\sim2-5$. The filled symbols present our measurements for galaxies at $z\sim2$ (blue triangles), $z\sim3$ (green circles), and $z\sim4-5$ (red pentagons). The magenta line is the best-fit linear $f_{\rm merger}^{\rm rel,\, cor}-\delta$ relation for the galaxies at $z\sim4-5$, and the magenta shaded region represents the $1.5\sigma$ typical scatter of the $z\sim4-5$ data points (Equation \ref{eq_delta_fmerg_relation}). The other symbols denote previous studies (green open circles: $z\sim3$ in \citealt{2016MNRAS.455.2363H}; green asterisks: $z\sim3$ in \citealt{2021ApJ...919...51M}; blue triangles: $z\sim2$ in \citealt{2023MNRAS.523.2422L}). All the galaxy merger fractions are shifted so that $f_{\rm merger}^{\rm rel,\, cor}$ is matched to $f_{\rm merger}^{\rm rel,\, cor}=0.2$ at $\delta\sim0$. Some data points are slightly shifted along the $x$ axis for clarity.}\label{fig_delta_fmerg}
\end{figure*}

% --------------------------------------------------
\subsection{Relation between Galaxy Merger Fraction and Galaxy Overdensity}\label{sec_relation}

We apply the super-resolution techniques to the galaxies at $z\sim2-5$, and estimate the galaxy merger fraction as a function of galaxy overdensity. In this study, we measure the {\it relative} galaxy merger fraction with respect to the field environment, corrected for the galaxy chance projection effect, $f_{\rm merger}^{\rm rel,\, cor}$. Figure \ref{fig_imgs} shows the super-resolution images of example galaxies at $z\sim2-5$. By selecting galaxy mergers with the combined images of the super-resolution techniques, we first calculate the galaxy merger fraction, $f_{\rm merger} = N_{\rm merger} / N$, where $N_{\rm merger}$ and $N$ are the numbers of galaxy mergers and sample galaxies, respectively. These numbers are obtained for the $z\sim2-5$ galaxies with $m_{\rm UV}<23.5$ selected in Section \ref{sec_selection}.

Next, we correct for the galaxy chance projection effect. Although we do not correct for $\mathit{TPR}$ and $\mathit{FPR}$ (Section \ref{sec_comparison_tech}), we need to take into account the chance projection effect of foreground/background sources. Our method to identify galaxy mergers relies on the projected separation, lacking redshift information for individual galaxy components. In higher $\delta$ regions, there is a higher likelihood of chance identifications of foreground/background sources near isolated galaxies. We correct for the chance projection effect by calculating 

% --------------------------------------------------
\begin{table}
  \tbl{Relative Galaxy Merger Fractions and Galaxy Merger Enhancement Ratios as a Function of Galaxy Overdensity}{%
  \begin{tabular}{ccc}
 \hline
 $\delta$ & $f_{\rm merger}^{\rm rel,\, cor}$ & $R_{\rm merger}$ \\
 (1) & (2) & (3) \\
  \hline
 \multicolumn{3}{c}{$z\sim2$} \\
$0.06$ & $0.200_{-0.004}^{+0.004}$ & $1.000_{-0.016}^{+0.016}$\\ 
$0.76$ & $0.206_{-0.009}^{+0.009}$ & $1.023_{-0.034}^{+0.034}$\\ 
\hline
 \multicolumn{3}{c}{$z\sim3$} \\
$0.06$ & $0.200_{-0.015}^{+0.015}$ & $1.000_{-0.091}^{+0.091}$\\ 
$0.94$ & $0.246_{-0.022}^{+0.025}$ & $1.288_{-0.138}^{+0.153}$\\ 
\hline
 \multicolumn{3}{c}{$z\sim4-5$} \\
$0.22$ & $0.200_{-0.005}^{+0.005}$ & $1.000_{-0.022}^{+0.022}$\\ 
$2.64$ & $0.277_{-0.044}^{+0.052}$ & $1.358_{-0.206}^{+0.240}$\\ 
$5.11$ & $0.359_{-0.107}^{+0.142}$ & $1.740_{-0.495}^{+0.661}$\\ 
\hline
\end{tabular}}\label{tab_delta_fmerg}
  \begin{tabnote}
 Note. (1) Average values of galaxy overdensity in each $\delta$ bin. (2) Relative galaxy merger fractions corrected for the chance projection effect. (3) Galaxy merger enhancement ratio, $f_{\rm merger}^{\rm rel,\, cor}/f_{\rm merger}^{\rm rel,\, cor}(\delta\sim0)$. 
  \end{tabnote}
\end{table}

\begin{equation}\label{eq_merger_frac_proj}
  f_{\rm merger}^{\rm col} = \frac{N_{\rm merger} - (1+\delta) N_{\rm proj}}{N}
\end{equation}

\noindent where $N_{\rm proj}$ is the expected number of chance projection sources in a galaxy sample. The expected number of chance projection sources per galaxy is estimated by multiplying the area of the galaxy merger search annulus $S=\pi(r_{\rm max}^2-r_{\rm min}^2)$ by the average surface number density of galaxies $n$ in the flux interval $0.25 f_1 \leq f < f_1$. To calculate $n$, we extract extended sources (i.e., galaxies) in the HSC-SSP UltraDeep COSMOS field by setting the {\tt extendedness\_value} flag to $1$ \citep{2018PASJ...70S...8A}. Then, we substitute $N_{\rm proj}=n\times S$ for Equation \ref{eq_merger_frac_proj}. 

Finally, we shift the galaxy merger fractions so that $f_{\rm merger}^{\rm col}$ nearest $\delta=0$ is aligned to a specific value, serving as the measurement for the field environment. We choose $f_{\rm merger}^{\rm col}=0.2$ as the galaxy merger fraction for the field environment, which is just for visualization purposes. The shifted $f_{\rm merger}^{\rm col}$ is referred to as the relative galaxy merger fraction, $f_{\rm merger}^{\rm rel,\, cor}$. 

Figure \ref{fig_delta_fmerg} presents $f_{\rm merger}^{\rm rel,\, cor}$ as a function of $\delta$. The $1\sigma$ uncertainties of $f_{\rm merger}^{\rm rel,\, col}$ are calculated by taking into account Poisson confidence limits \citep{1986ApJ...303..336G} on the numbers of the sources. For $z\sim4-5$, the low, intermediate, and high overdensity bins are defined as $\delta_{\rm low}<2$, $\delta_{\rm mid}=2-4$, and $\delta_{\rm high}=4-10$, respectively. For $z\sim2-3$, the samples are divided by a small $\delta$ value of $\delta=0.5$, i.e., $\delta_{\rm low}<0.5$, $\delta_{\rm mid}=0.5-4$ due to relatively small $\delta$ coverages. Table \ref{tab_delta_fmerg} lists $f_{\rm merger}^{\rm rel,\, cor}$ in each galaxy overdensity bin. We also plot previous studies for $z\sim2-3$ galaxy protoclusters whose $f_{\rm merger}$ and $\delta$ are available: \citet{2016MNRAS.455.2363H} and \citet{2021ApJ...919...51M} for the SSA22 galaxy protocluster at $z=3.09$ and \citet{2023MNRAS.523.2422L} for two galaxy protoclusters of BOSS1244 and BOSS1542 at $z=2.24$. The galaxy merger fractions for the previous studies are shifted in the same way as for our measurements. 

Despite the relatively narrow $\delta$ coverage, our $f_{\rm merger}^{\rm rel,\, col}$ measurements at $z\sim3$ validate the results of the previous studies suggesting that $f_{\rm merger}^{\rm rel,\, col}$ increases with increasing $\delta$ at $z\sim2-3$ \citep{2016MNRAS.455.2363H, 2023MNRAS.523.2422L}. In addition, we find a similar increasing trend of $f_{\rm merger}^{\rm rel,\, col}$ at higher redshifts of $z\sim4-5$. To our knowledge, this represents the highest-$z$ evidence  that galaxy mergers occur more frequently in denser regions, identified with a statistical galaxy sample. The combination of the wide-area HSC-SSP data and the super-resolution techniques has allowed us to reveal the relation between $f_{\rm merger}^{\rm rel,\, col}$ and $\delta$ at the highest redshifts. Interestingly, $f_{\rm merger}^{\rm rel, col}$ at $z\sim4-5$ appears to increase almost linearly with increasing $\delta$. The $f_{\rm merger}^{\rm rel,\, col}$ at $z\sim4-5$ is fitted with the following linear function, 

% ---------- Figure 
\begin{figure*}
 \begin{center}
 \includegraphics[width=170mm]{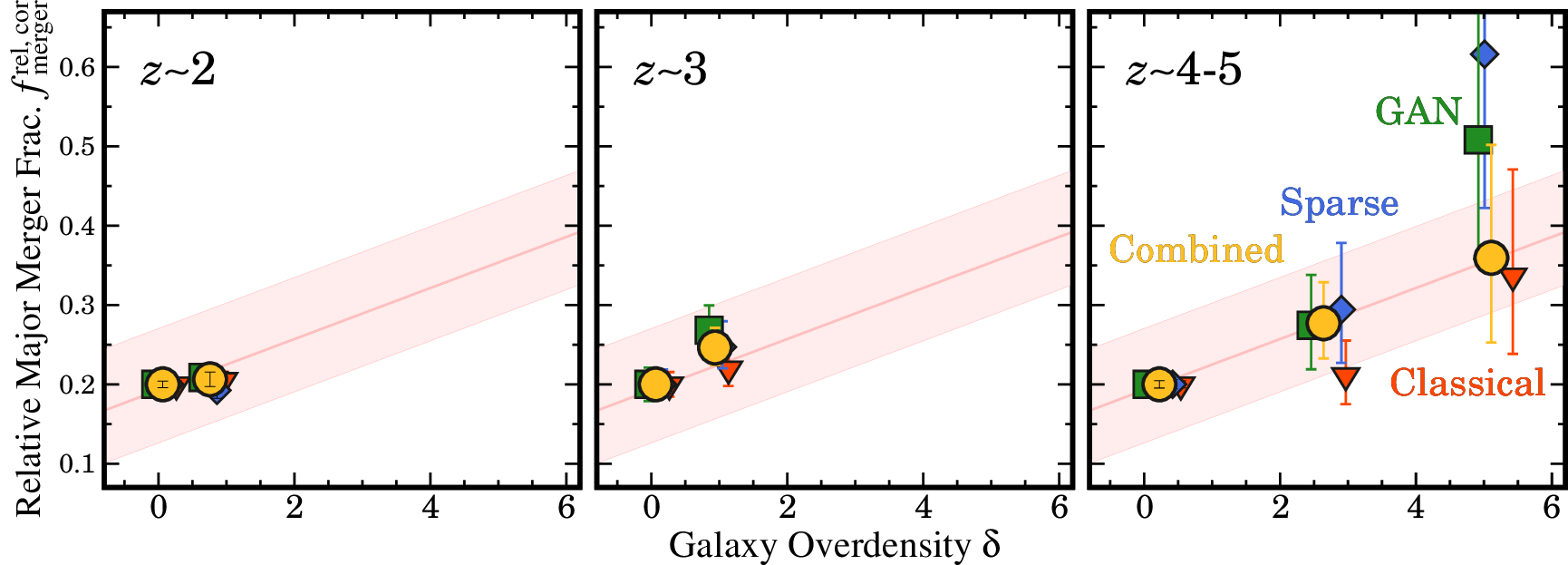}
 \end{center}
   \caption{Same as Figure \ref{fig_delta_fmerg}, but for the relative galaxy merger fractions calculated in various types of super-resolution techniques (red inverse-triangles; classical RL PSF  deconvolution; blue diamonds: sparse modeling; green squares: GAN; yellow circles: combined). The left, middle, and right panels indicate the relative galaxy merger fractions for galaxies at $z\sim2$, $z\sim3$, and $z\sim4-5$, respectively.}\label{fig_delta_fmerg_methods}
\end{figure*}

\begin{figure*}
 \begin{center}
 \includegraphics[width=170mm]{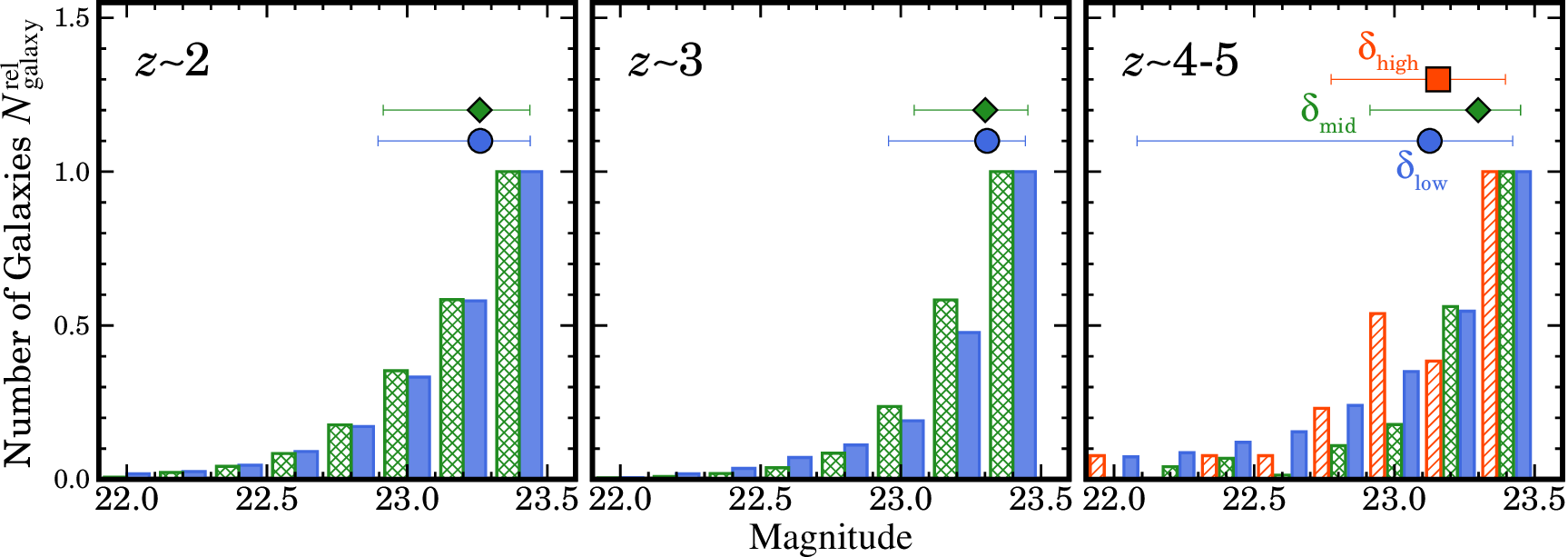}
 \end{center}
   \caption{Number of galaxies as a function of magnitude. The blue filled, green cross-hatched, and red single-hatched histograms represent the low, intermediate, and high $\delta$ bins, respectively. The peaks of the galaxy numbers are matched. The median values of magnitudes in each $\delta$ bin are shown with the same color coding as for the histograms (blue circles: the low $\delta$ bin; green diamonds: the intermediate $\delta$ bin; red square: the high $\delta$ bin). The error bars indicate the 16th and 84th percentiles of the magnitude distribution. The left, middle, and right panels show  galaxies at $z\sim2$, $z\sim3$, and $z\sim4-5$, respectively.}\label{fig_hist_mag_delta}
\end{figure*}
% ---------- Figure 

% ---------- Table 
\begin{longtable}{*{6}{c}}
\caption{Probabilities of the Kolmogorov-Smirnov Tests for the Magnitude Distributions in the $\delta$ bins$^*$}\label{tab_ks_test}
 \hline
  \multicolumn{1}{c}{$z\sim2$} & \multicolumn{1}{c}{$z\sim3$} & \multicolumn{4}{c}{$z\sim4-5$} \\
 \cmidrule(lr){1-1} \cmidrule(lr){2-2} \cmidrule(lr){3-6}
 $(\delta_{\rm low}, \delta_{\rm mid})$ & $(\delta_{\rm low}, \delta_{\rm mid})$ & $(\delta_{\rm mid}, \delta_{\rm high})$ & $(\delta_{\rm low}, \delta_{\rm high})$ & $(\delta_{\rm low}, \delta_{\rm mid})$ & $(\delta_{\rm low}, \delta_{\rm mid})^\dagger$ \\
 \endhead
 \hline
 $0.500$ & $0.329$ & $0.247$ & $0.087$ & $(2.465\times10^{-9})$ & $0.514$ \\
  \hline
\multicolumn{6}{l}{$^*$ The value in the parenthesis is a $p$-value smaller than the significance level of $5$\%.} \\
\multicolumn{6}{l}{$\dagger$ The $p$-value of the KS test for distributions in a narrow magnitude range of $m=23-23.5$.} \\
\end{longtable}
% ---------- Table 

\begin{equation}\label{eq_delta_fmerg_relation}
f_{\rm merger}^{\rm rel,\, col}(\delta) = 0.032\times\delta + 0.19 (^{+0.08}_{-0.07})
\end{equation}

\noindent The uncertainty in the parenthesis in Equation \ref{eq_delta_fmerg_relation} represents the $1.5\sigma$ typical scatter of the $z\sim4-5$ data points (the magenta shaded region in Figure \ref{fig_delta_fmerg}). The reason for the slight difference between the intercept of Equation \ref{eq_delta_fmerg_relation} (i.e., 0.19) and $f_{\rm merger}^{\rm rel,\, col}=0.2$ is that the average $\delta$ value within the $\delta$ bin of the field environment is slightly offset from $\delta=0$ (Table \ref{tab_delta_fmerg}). We find that $f_{\rm merger}^{\rm rel,\, col}$ measured in this study and the literature, including \citet{2021ApJ...919...51M} who suggest that $f_{\rm merger}^{\rm rel,\, col}$ does not correlate with $\delta$, broadly follows the linear $f_{\rm merger}^{\rm rel,\, col}-\delta$ relation within the uncertainty. The lack of $f_{\rm merger}^{\rm rel,\, col}-\delta$ correlation of \citet{2021ApJ...919...51M} might result from a small sample of galaxies ($N_{\rm galaxy}=63$). 

On the contrary, the $f_{\rm merger}^{\rm rel,\, col}-\delta$ relation for our $z\sim2$ sample is weak or almost flat. The weak or missing $f_{\rm merger}^{\rm rel,\, col}-\delta$ relation at $z\sim2$ is perhaps due to the narrower area of the CLAUDS survey than the HSC-SSP survey (Table \ref{tab_sample}). A wider survey for $z\sim2$ might be required to confirm the $f_{\rm merger}^{\rm rel,\, col}-\delta$ relation at $z\sim2$. 

In Section \ref{sec_comparison_prev}, we conduct detailed comparisons between $f_{\rm merger}^{\rm rel,\, col}$ measured in this study and the literature at $z\sim0-1$ over a wide $\delta$ range of $\delta\sim0-20$ to investigate whether the $f_{\rm merger}^{\rm rel,\, col}-\delta$ relation evolves with redshift or not.  

% --------------------------------------------------
\subsection{Is the Galaxy Merger-Overdensity Relation Real?}\label{sec_real}

We examine whether the $f_{\rm merger}^{\rm rel,\, col}-\delta$ trend is real or not. Our $f_{\rm merger}^{\rm rel,\, col}$ results are based on the application of the super-resolution techniques that computationally enhance the spatial resolution of the seeing-limited HSC images. The $f_{\rm merger}^{\rm rel,\, col}-\delta$ trend might be artificially produced during the image processing analysis because of some possible systematics. Here we check whether the $f_{\rm merger}^{\rm rel,\, col}-\delta$ relation is affected by differences in (1) the super-resolution techniques, (2) the magnitude between the $\delta$ bins, and (3) the depth of the survey fields.  

\begin{enumerate}
 \item {\it Difference in the super-resolution techniques.} --- We test whether the difference in the super-resolution techniques affects the $f_{\rm merger}^{\rm rel,\, col}-\delta$ relation. Figure \ref{fig_delta_fmerg_methods} presents $f_{\rm merger}^{\rm rel,\, col}$ obtained with the three super-resolution techniques. At $z\sim2$ and $z\sim3$, we confirm that the classical RL PSF deconvolution, sparse modeling, and GAN-based $f_{\rm merger}^{\rm rel,\, col}$ are in good agreements with those of the combined images. Similarly, at $z\sim4-5$, we also find that $f_{\rm merger}^{\rm rel,\, col}$ measured in the three super-resolution techniques follows the $f_{\rm merger}^{\rm rel,\, col}-\delta$ relation given the uncertainties. These agreements indicate that the $f_{\rm merger}^{\rm rel,\, col}-\delta$ relation does not strongly depend on the super-resolution techniques.

\begin{figure}
 \begin{center}
 \includegraphics[width=80mm]{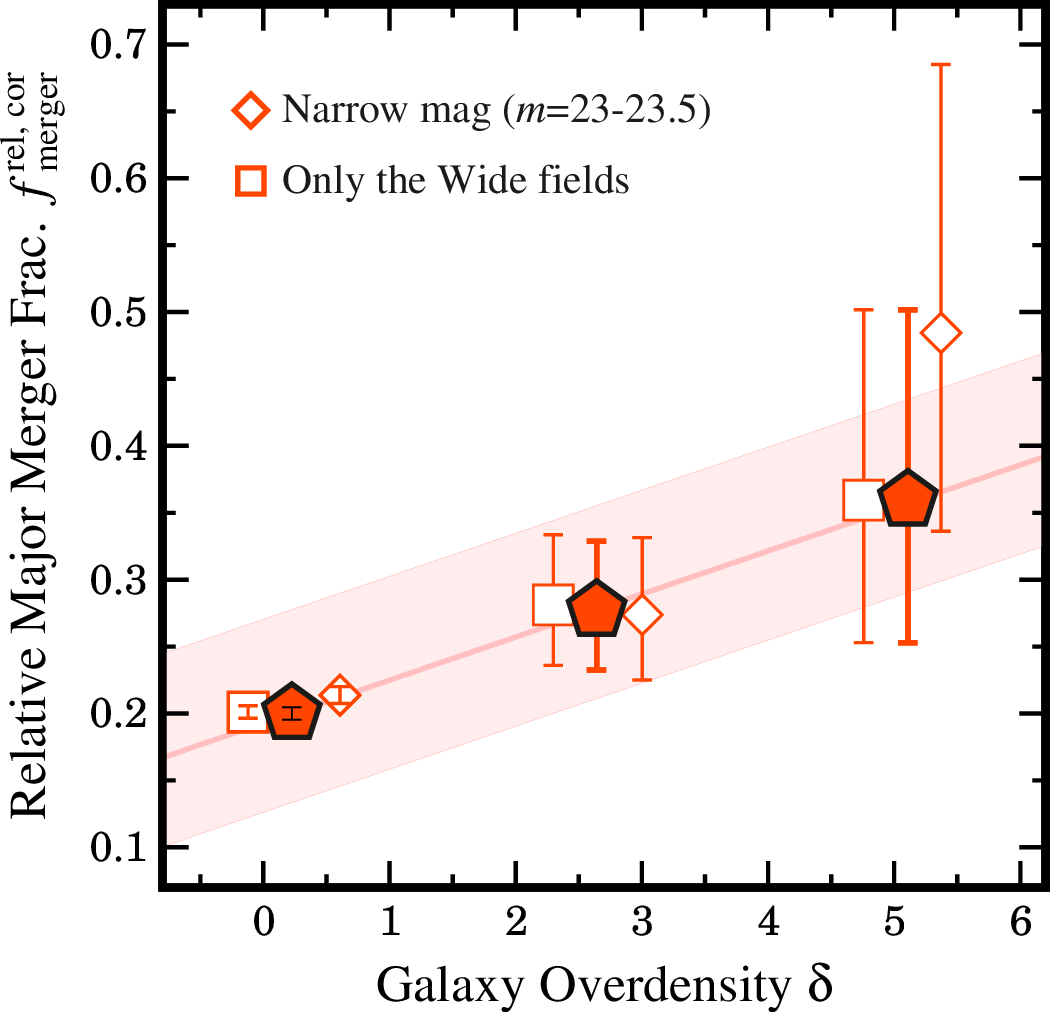}
 \end{center}
   \caption{Same as Figure \ref{fig_delta_fmerg}, but for the relative galaxy merger fractions at $z\sim4-5$ calculated in three different criteria to select sample galaxies (filled pentagons; the fiducial sample, which is the same as the one in Figure \ref{fig_delta_fmerg}; open diamonds: a sample of galaxies selected in a narrow magnitude range of $m=23-23.5$; open squares: a sample of galaxies located in the HSC Wide fields).}\label{fig_delta_fmerg_narrow_mag}
\end{figure}

 \item {\it Difference in the magnitude between the $\delta$ bins.} --- We examine the difference in magnitude between the $\delta$ bins. Typically, galaxy mergers are identified with a higher $\mathit{TPR}$ value for brighter galaxies (Figure \ref{fig_tpr_fpr}). If bright galaxies are located selectively in high $\delta$ regions, it might potentially result in an artificial increasing $f_{\rm merger}^{\rm rel,\, col}-\delta$ trend due to the high $\mathit{TPR}$ value. Figure \ref{fig_hist_mag_delta} shows the histograms of magnitude between the $\delta$ bins and the median values of magnitudes. Within the $1\sigma$ uncertainties, the average magnitudes are comparable between the $\delta$ bins. In addition, the magnitude distribution appears similar between the $\delta$ bins. To quantify the difference in the magnitude distribution, we perform the Kolmogorov-Smirnov (KS) tests. Table \ref{tab_ks_test} summarizes the $p$-values of the KS tests. We find that the $p$-values exceed the commonly-used significance level of $5$\% for almost all the distributions. The KS tests cannot reject a null hypothesis that two samples are drawn from a statistically identical distribution. Among the distributions, one combination of the low- and mid-$\delta$ bins, ($\delta_{\rm low}, \delta_{\rm mid}$), at $z\sim4-5$ shows $p<5$\%. To reduce the magnitude difference, we measure $f_{\rm merger}^{\rm rel,\, col}-\delta$ in a relatively narrow magnitude range of $m=23-23.5$. With this narrow $m$ range, the KS test for the ($\delta_{\rm low}, \delta_{\rm mid}$) combination shows $p>5$\% (Table \ref{tab_ks_test}). We find that the new $f_{\rm merger}^{\rm rel,\, col}$ values measured in the narrow $m$ range still follow the $f_{\rm merger}^{\rm rel,\, col}-\delta$ relation given the uncertainties (Figure \ref{fig_delta_fmerg_narrow_mag}). The comparable median magnitudes and the $p$-values suggest that the difference in magnitude between the $\delta$ bins does not significantly influence the $f_{\rm merger}^{\rm rel,\, col}-\delta$ relation. 

 \item {\it Difference in the depth of the survey fields.} --- We investigate whether the $f_{\rm merger}^{\rm rel,\, col}-\delta$ relation is affected or not by the depths of the different survey fields. Rare high-$\delta$ regions are typically identified in the HSC Wide fields whose $\mathit{FPR}$ is higher than that in the HSC UltraDeep and Deep fields (Figure \ref{fig_tpr_fpr}). There is a possibility that $f_{\rm merger}^{\rm rel,\, col}$ is systematically higher in the higher $\delta$ bins due to the high $\mathit{FPR}$ value of the HSC Wide fields. To examine the effect, we measure $f_{\rm merger}^{\rm rel,\, col}$ using only the galaxy sample in the HSC Wide fields (Figure \ref{fig_delta_fmerg_narrow_mag}).\footnote{We are not able to obtain statistically meaningful results of the $f_{\rm merger}^{\rm rel,\, col}-\delta$ relation using only the Deep/UltraDeep field data.} We find that $f_{\rm merger}^{\rm rel,\, col}$ obtained in only the HSC Wide fields agrees with the $f_{\rm merger}^{\rm rel,\, col}-\delta$ relation within the uncertainty. The agreement suggests that the $f_{\rm merger}^{\rm rel,\, col}-\delta$ relation is not a result of the difference in the depth of the survey fields.

\end{enumerate}

\noindent These tests indicate that the $f_{\rm merger}^{\rm rel,\, col}-\delta$ relation is unlikely to be caused by the differences in the super-resolution techniques, magnitudes between the $\delta$ bins, or the depth of the survey fields. Therefore, we conclude that the $f_{\rm merger}^{\rm rel,\, col}-\delta$ relation is real. 

% --------------------------------------------------
% --------------------------------------------------
% --------------------------------------------------
\section{Discussion}\label{sec_discussion}

% --------------------------------------------------
\subsection{Comparisons with Previous Studies at Low Redshifts; Redshift Evolution of the Galaxy Merger Fraction-Galaxy Overdensity Relation}\label{sec_comparison_prev}

We compare $f_{\rm merger}^{\rm rel,\, col}$ in this study with results in previous studies at low redshifts to investigate whether the $f_{\rm merger}^{\rm rel,\, col}-\delta$ relation evolves or remains unchanged over cosmic time. As described below, the $f_{\rm merger}^{\rm rel,\, col}-\delta$ relation does not strongly evolve in a redshift range of $z\sim2-5$, while the galaxy merger fractions are possibly suppressed in massive galaxy clusters at $z\lesssim1$. In Section \ref{sec_relation}, we have found that the galaxy merger fraction broadly follows the linear $f_{\rm merger}^{\rm rel,\, col}-\delta$ relation at high redshifts of $z\sim2-5$. Here we include previous studies at $z\lesssim1$ and a high galaxy overdensity of $\delta\gtrsim5$ as comparison samples. Table \ref{tab_observational} summarizes observational studies examining the relation between galaxy mergers and galaxy environments. Appendix provides details about the previous studies, e.g., galaxy samples and methods to estimate the galaxy overdensity for each work (Section \ref{sec_appendix}). For the plot, we select previous studies whose galaxy merger fractions are available for both field and overdense environments and have been measured in the galaxy close-pair method.  Although the methods to identify galaxy mergers are slightly different from ours (e.g., the separation between galaxy merger components), the comparisons yield insights into understanding the general trend of the $f_{\rm merger}^{\rm rel,\, col}-\delta$ relation. 

\begin{figure*}
 \begin{center}
 \includegraphics[width=140mm]{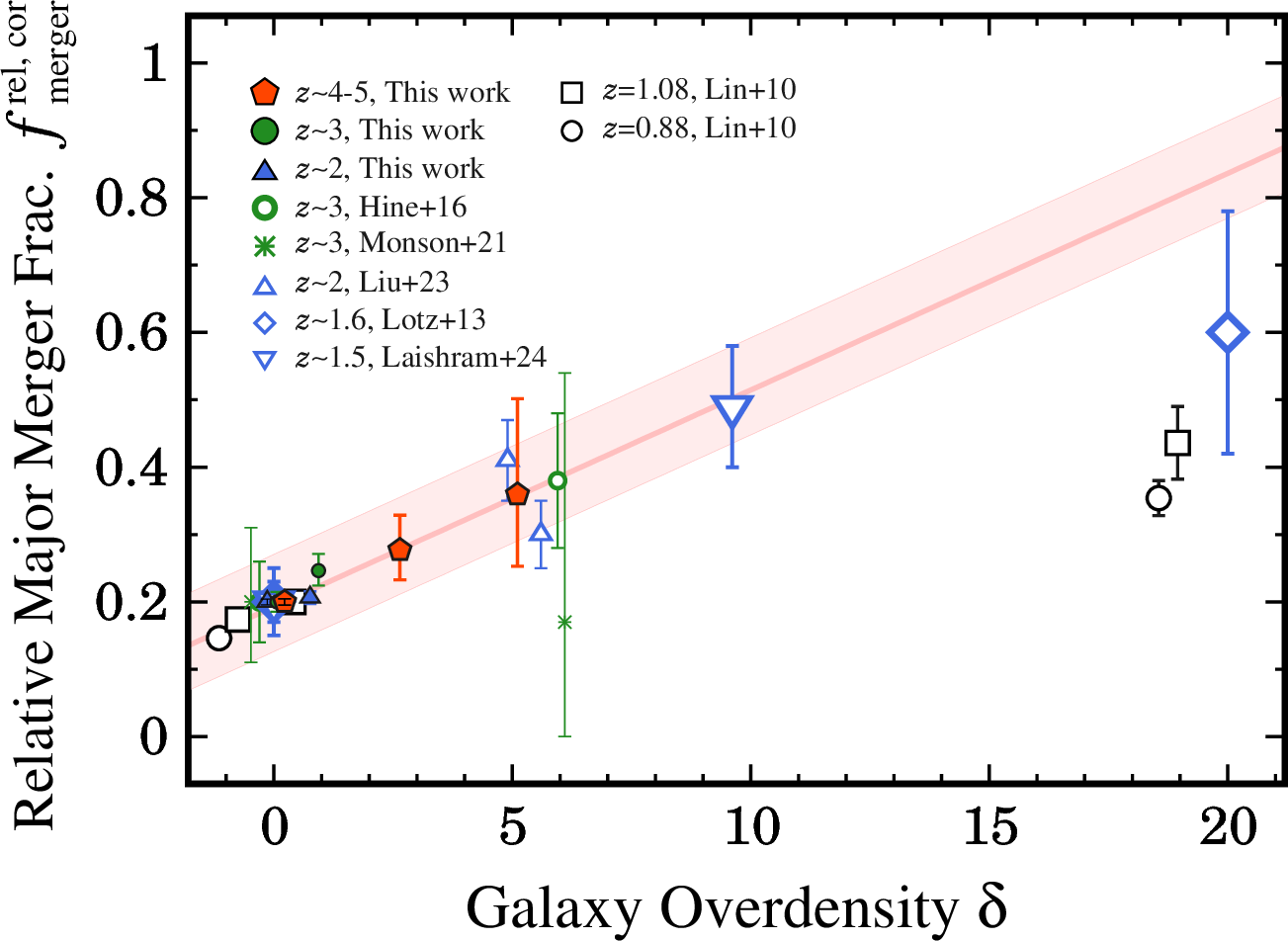}
 \end{center}
   \caption{Same as Figure \ref{fig_delta_fmerg}, but extending the range of galaxy overdensity to $\delta\sim20$ and including previous studies at $z<1$. The blue diamonds represent \citet{2013ApJ...773..154L}. The blue inverse triangles indicate \citet{2024ApJ...964L..33L}. The black squares and black circles denote measurements in \citet{2010ApJ...718.1158L} for galaxies at average redshifts of $z=1.08$ and $z=0.88$, respectively.}\label{fig_delta_fmerg_comparison}
\end{figure*}

Figure \ref{fig_delta_fmerg_comparison} presents $f_{\rm merger}^{\rm rel,\, col}$ as a function of $\delta$ including three previous studies of \citet{2013ApJ...773..154L}, \citet{2024ApJ...964L..33L}, and \citet{2010ApJ...718.1158L}. \citet{2013ApJ...773..154L} have estimated the frequency of double nuclei or close companions in a $\delta\sim20$ massive protocluster at $z\sim1.6$. \citet{2024ApJ...964L..33L} have investigated the morphological disturbance of galaxies in extensive filamentary structures at $z\sim1.5$. \citet{2010ApJ...718.1158L} have used a spectroscopic sample of $\sim20,000$ galaxies, and have evaluated the galaxy merger fraction at two average redshifts of $z=1.08$ and $z=0.88$. These studies suggest that $f_{\rm merger}^{\rm rel,\, col}$ increases with increasing $\delta$. At $\delta\sim10$ and $20$, we find that the galaxy merger fractions at the high redshift of $z\sim1.5-1.6$ follows the extrapolation of the $f_{\rm merger}^{\rm rel,\, col}-\delta$ relation within the uncertainty. However, the two $f_{\rm merger}$ measurements of \citet{2010ApJ...718.1158L} at $z\sim1$ show significant deviations from the $f_{\rm merger}^{\rm rel,\, col}-\delta$ relation. The deviation from the $f_{\rm merger}^{\rm rel,\, col}-\delta$ relation appears to be more significant at lower redshifts. 

In addition to $f_{\rm merger}^{\rm rel,\, col}$, we measure the enhancement ratio of the galaxy merger fraction, $R_{\rm merger}$, to reduce the impact of the redshift evolution of $f_{\rm merger}$ in the field environment (e.g., \citealt{2011ApJ...742..103L, 2016ApJ...830...89M, 2017MNRAS.470.3507M, 2018MNRAS.475.1549M, 2022ApJ...940..168C}). The enhancement ratio of the galaxy merger fraction is defined as, 

\begin{equation}\label{eq_merger_enhancement}
 R_{\rm merger} = \frac{f_{\rm merger}^{\rm rel,\, col}}{f_{\rm merger}^{\rm rel,\, col}(\delta\sim0)}, 
\end{equation}

\noindent where $f_{\rm merger}^{\rm rel,\, col}(\delta\sim0)$ indicates the galaxy merger fraction at $\delta\sim0$. 

\begin{figure*}
 \begin{center}
 \includegraphics[width=145mm]{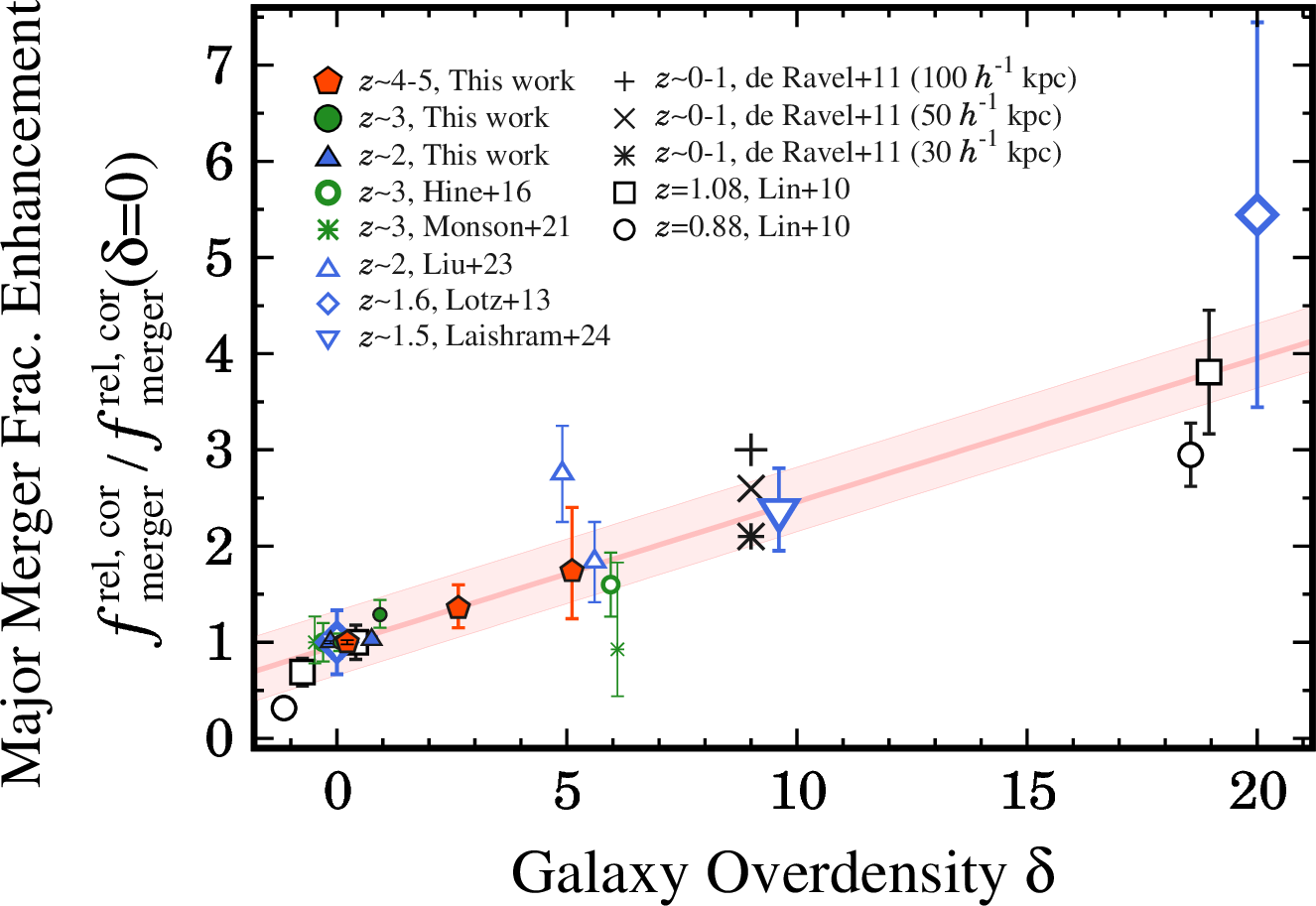}
 \end{center}
   \caption{Enhancement ratio of the galaxy merger fraction, $R_{\rm merger}=f_{\rm merger}^{\rm rel,\, cor}/f_{\rm merger}^{\rm rel,\, cor}(\delta\sim0)$, as a function of galaxy overdensity. The black plus mark, cross, and asterisk indicate \citet{2011arXiv1104.5470D}'s $R_{\rm merger}$ calculated in three different selection criteria of galaxy mergers with galaxy-galaxy separations of $d=100$, $50$, and $30\, h^{-1}$ kpc, respectively. The definitions of the line and the other symbols are the same as those in Figures \ref{fig_delta_fmerg} and \ref{fig_delta_fmerg_comparison}.}\label{fig_delta_fmerg_enhancement}
\end{figure*}

Figure \ref{fig_delta_fmerg_enhancement} presents the enhancement ratio of the galaxy merger fraction as a function of $\delta$. From Equation \ref{eq_delta_fmerg_relation}, the linear $R_{\rm merger}-\delta$ relation (the magenta line and magenta shaded region) is calculated as, 

\begin{equation}\label{eq_delta_r_relation}
R_{\rm merger}(\delta) = 0.15\times\delta + 0.97 (^{+0.36}_{-0.31}). 
\end{equation}

\begin{longtable}{*{5}{c}}
\caption{Observational studies investigating the relation between $f_{\rm merger}$ and $\delta$ with $>10$ sample galaxies}\label{tab_observational}
 \hline
 Reference & Redshift & $N_{\rm galaxy}$ & Method & $f_{\rm merger}-\delta$ related? \\
 (1) & (2) & (3) & (4) & (5) \\
 \endhead
 \hline
  This work & $2-5$ & 32,187 & Super-resolution, Pair & $+$ \\ 
  \citet{2016MNRAS.455.2363H} & $3.1$ & 109 & Visual & $+$ \\ 
  \citet{2021ApJ...919...51M} & $3.1$ & 63 & $n$, $GM_{20}$, Visual & No \\ 
  \citet{2023AA...670A..58M} & $1.3-2.8$ & $271$ & Visual & $+$ \\ 
  \citet{2022arXiv221205984M} & $2-2.5$ & $1,344^a$ & Pair & No \\ 
  \citet{2023MNRAS.523.2422L} & $2.24$ & 577 & Pair & $+$ \\ 
  \citet{2017ApJ...843..126D} & $1.59-1.71$ & 146 & Visual & No or `$-$' \\ 
  \citet{2018MNRAS.479..703C} & $1.99$ & 11 & Visual & $+$ \\ 
  \citet{2020ApJ...899...85S} & $1.2-1.8$ & 498 & $A, GM_{20}$, Pair & $+$ \\ 
  \citet{2013ApJ...773..154L} & $1.62$ & $144$ & Pair & $+$ \\ 
  \citet{2019ApJ...874...63W} & $1.6$ \& $1.9$ & 949 & Visual, Pair & $+$ \\ 
  \citet{2024ApJ...964L..33L} & $1.5$ & 368 & $GM_{20}$ & $+$ \\ 
  \citet{2010ApJ...718.1158L} & $0.75-1.2$ & $\sim20,000$ & Pair & $+$ \\ 
  \citet{2013ApJ...762...43K} & $0.2-1$ & 3,667 & Pair & $+$ \\ 
  \citet{2011arXiv1104.5470D} & $0.2-1$ & 10,644 & Pair & $+$ \\ 
  \citet{2011ApJ...736...38K} & $0.9$ & 201 & $GM_{20}$, Visual & $+$ \\ 
  \citet{2020ApJ...899...64A} & $0.9$ & 101 & $GM_{20}$, Visual & Groups \\ 
  \citet{2019AA...630A..57P} & $0.84$ & 490 & $A$, $GM_{20}$, Visual & No \\ 
  \citet{1999ApJ...520L..95V} & $0.83$ & 81 & Visual & `$+$', outskirts \\ 
  \citet{2023AA...679A.142O} & $0.01-0.35$ & 302,148 & DL & $-$ \\ 
  \citet{2024arXiv240218520S} & $0.1-0.15$ & 23,855 & DL, $GM_{20}$ & $-$ \\ 
  \citet{2008MNRAS.388.1537M} & $<0.12$ & 5,376 & Pair, Residuals & Groups \\ 
  \citet{2009MNRAS.396.2003L} & $0.03-0.12$ & 515 & Pair, Residuals, $A$ & $+$ \\ 
  \citet{2012AA...539A..46A} & $0-0.1$ & 660 & Pair & Low-$\delta$ groups \\ 
  \citet{2009MNRAS.399.1157P} & $0.01-0.1$ & $\cdots$ & Pair & Intermediate \\ 
  \citet{2010MNRAS.407.1514E} & $<0.1$ & 5,784 & Pair, $A$ & See appendix \\ 
  \citet{2010MNRAS.401.1552D} & $0.005-0.1$ & 3,003 & Visual & `$+$' or weak \\ 
  \citet{2006MNRAS.373..167M} & $\sim0$ & 680 & Visual & Intermediate \\ 
 \hline
\multicolumn{5}{l}{(1) Reference. (2) Redshift of galaxies. (3) Number of galaxies. (4) Method to select galaxy mergers. } \\
\multicolumn{5}{l}{Pair: close-pair; Visual: visual inspection; $n$: S\'ersic index; Residual: analyzing residual images after } \\
\multicolumn{5}{l}{the radial profile fitting of galaxies; $A$: Asymmetry; $GM_{20}$: Gini coefficient and } \\
\multicolumn{5}{l}{the second-order moment of the brightest $20$\% of galaxy pixel; DL: Deep learning. } \\
\multicolumn{5}{l}{(5) Existence of the $f_{\rm merger}-\delta$ relation. ``+": $f_{\rm merger}$ increases with increasing $\delta$; } \\
\multicolumn{5}{l}{``$-$": $f_{\rm merger}$ decreases with increasing $\delta$; ``No": $f_{\rm merger}$ is not related to $\delta$. The other words } \\
\multicolumn{5}{l}{such as `Groups' and `Intermediate' indicate environments with the highest galaxy merger fraction. } \\
\multicolumn{5}{l}{See also Appendix (Section \ref{sec_appendix}) for the brief explanation of each work. } \\
\multicolumn{5}{l}{$^a$ Updated from the number in the original paper (R. Momose, private communication). } \\
\end{longtable}

\noindent This $R_{\rm merger}$ function is roughly consistent with the environmental dependence predicted by an N-body simulation of \cite{2009MNRAS.394.1825F} suggesting that the halo merger rate at $z\sim2$ is $\sim1.2-1.3$ times enhanced at $\delta=3$ compared to the field environment. In Figure \ref{fig_delta_fmerg_enhancement}, we add three enhancement ratios of the galaxy merger fraction measured using $10,644$ galaxies at $z\sim0-1$ in \citet{2011arXiv1104.5470D}. As in the figure of $f_{\rm merger}^{\rm rel,\, col}$ (Figure \ref{fig_delta_fmerg_comparison}), almost all of the $f_{\rm merger}^{\rm rel,\, col}$ measurements broadly agree with the $R_{\rm merger}-\delta$ relation at $\delta\sim0-10$ although some data points slightly deviate from the error region (e.g., a data point in \citealt{2023MNRAS.523.2422L}). Particularly, at a high galaxy overdensity of $\delta\sim20$, $R_{\rm merger}$ at $z\sim1-2$ shows a good agreement with the $R_{\rm merger}-\delta$ relation. However, $R_{\rm merger}$ at $z<1$ might still be below the $R_{\rm merger}-\delta$ relation. 

In summary, the environmental dependence of galaxy mergers does not strongly evolve over cosmic time of $z\sim2-5$ according to the diagrams of $f_{\rm merger}^{\rm rel,\, col}-\delta$ and $R_{\rm merger}-\delta$. However, in massive galaxy clusters at $z\lesssim1$, the galaxy merger fraction appears to be suppressed compared to the $f_{\rm merger}^{\rm rel,\, col}-\delta$ relation, and possibly even below the $R_{\rm merger}-\delta$ relation. This trend is consistent with predictions of theoretical studies with N-body simulations (e.g., \citealt{2001ApJ...546..223G}), and semi-analytic models (e.g., \citealt{2012ApJ...754...26J}). These theoretical studies predict that $f_{\rm merger}$ tends to be high in the galaxy group/cluster environments at $z>1$, while $f_{\rm merger}$ rapidly decreases in massive galaxy clusters at $z<1$. The decrease in $f_{\rm merger}$ is attributed to high-velocity dispersions of massive galaxy clusters, where high-velocity galaxy encounters hinder galaxy mergers and interactions \citep{1980ComAp...8..177O, 1997ApJ...481...83M}. In addition, we have found that this $f_{\rm merger}$ suppression at high-$\delta$ takes place at $z\sim1$, which is roughly consistent with the predicted redshift when $f_{\rm merger}$ is suppressed in simulations \citep{2001ApJ...546..223G}. Our $f_{\rm merger}-\delta$ and $R_{\rm merger}-\delta$ relations might be useful and additional constraints on cosmological simulations for galaxy overdense regions at high-$z$ (e.g., \citealt{2013MNRAS.429..323S, 2013ApJ...779..127C, 2015MNRAS.452.2528M, 2016MNRAS.456.1924C, 2017MNRAS.470.4186B, 2017MNRAS.471.1088B, 2018MNRAS.479.5385H, 2018MNRAS.480.2898C, 2019MNRAS.483.3336T, 2022MNRAS.509.4037Y, 2024arXiv240208911N}). 

% --------------------------------------------------
\subsection{Role of Galaxy Mergers in Galaxy Protoclusters}\label{sec_role_galaxy_mergers}

We discuss the role of galaxy mergers in the environmental dependence of galaxy populations. As presented in Section \ref{sec_comparison_prev} and Figure \ref{fig_delta_fmerg_enhancement}, we have found that the enhancement ratio of the galaxy merger fraction, $R_{\rm merger}$, can be roughly predicted from the galaxy overdensity at least in the redshift range of $z\sim2-5$. Using the $R_{\rm merger}-\delta$ relation, we are able to evaluate the contribution level of galaxy mergers to establishing various types of the high-$z$ environmental dependence. As a demonstration, we focus on one of the widely studied populations for the environmental dependences: AGN. Because galaxy mergers and interactions are thought to trigger the AGN activity (e.g., \citealt{1985AJ.....90..708K, 1988ApJ...325...74S, 2005Natur.433..604D, 2008ApJ...674...80U, 2010ApJ...716L.125K, 2011ApJ...743....2S, 2011MNRAS.418.2043E, 2014MNRAS.441.1297S, 2018PASJ...70S..37G}; see also a summary table of \citealt{2023OJAp....6E..34V}), the AGN fraction, $f_{\rm AGN}$, is expected to be enhanced in high-$z$ galaxy protoclusters, as found in the literature (e.g., \citealt{2009ApJ...691..687L, 2010MNRAS.407..846D, 2013ApJ...765...87L, 2023PASJ...75.1246H, 2024ApJ...967...65T}). In this analysis, we use a summary table of \citet{2019ApJ...874...54M}, which compiles measurements of $f_{\rm AGN}$ for galaxy protoclusters and the field environment at $z\sim2-3$. 

As in the equation of $R_{\rm merger}$ (Equation \ref{eq_merger_enhancement}), the enhancement ratio of $f_{\rm AGN}$ is defined as 

\begin{equation}\label{eq_agn_enhancement}
R_{\rm AGN} = \frac{f_{\rm AGN}}{f_{\rm AGN}(\delta\sim0)}, 
\end{equation}

\begin{figure}
 \begin{center}
 \includegraphics[width=80mm]{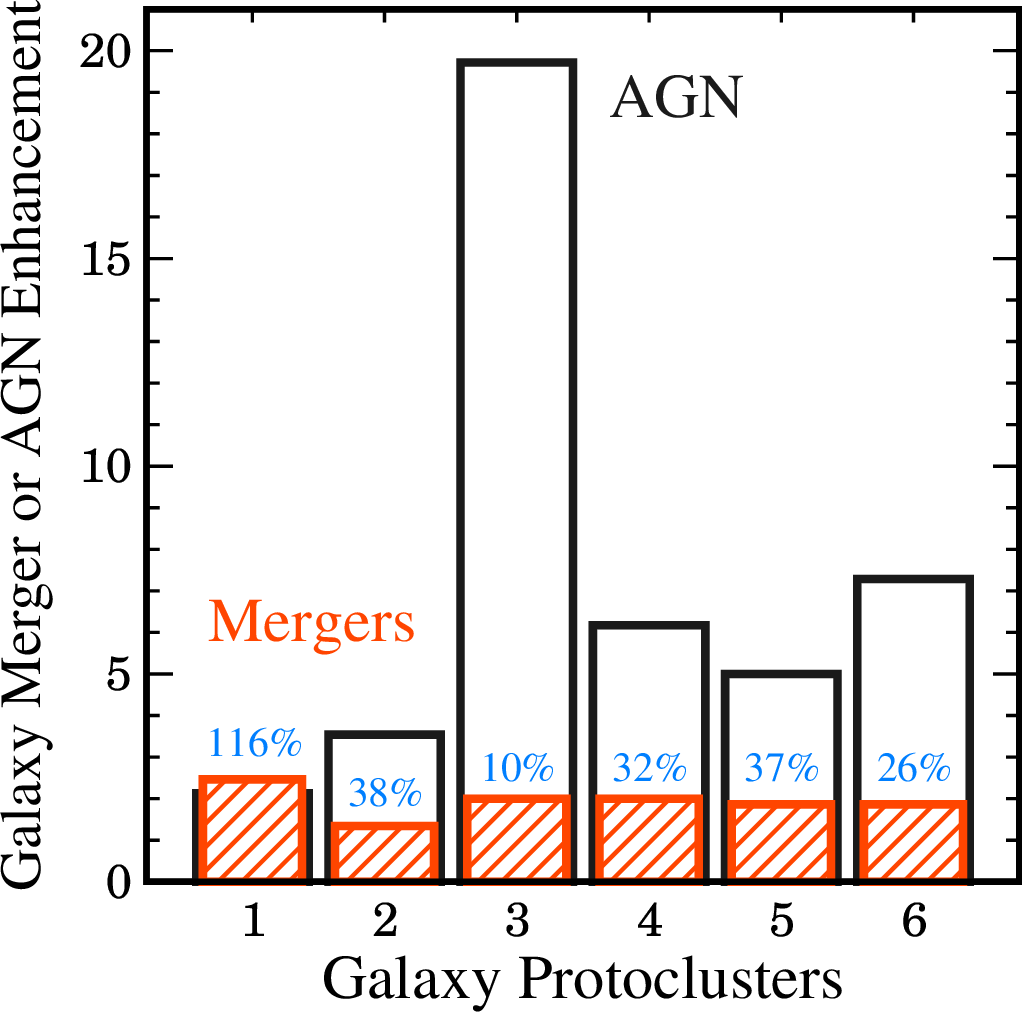}
 \end{center}
   \caption{Enhancement ratios of the galaxy merger fraction $R_{\rm merger}$ (red hatched histogram) and the AGN fraction $R_{\rm AGN}$ (black open histogram) relative to the field environment for $z\sim2-3$ galaxy protoclusters (1: Cl 0218.3-0510 at $z=1.6$; 2: 2QZ 1004+00 at $z=2.23$; 3: BX/MD galaxies in QSO HS 1700+643 at $z=2.3$; 4: Ly$\alpha$ emitting galaxies in QSO HS 1700+643 at $z=2.3$; 5: Lyman break galaxies in SSA22 at $z=3.09$; 6: Ly$\alpha$ emitting galaxies in SSA22 at $z=3.09$). See \citet{2019ApJ...874...54M} for details of the galaxy protoclusters and the method to calculate $f_{\rm AGN}$. The values above the red hatched histogram indicate the fractional contributions, $R_{\rm merger}/R_{\rm AGN}$.}\label{fig_agn_merger_enhancement}
\end{figure}

\noindent where $f_{\rm AGN}(\delta\sim0)$ is the AGN fraction in the field environment. Using the $f_{\rm AGN}$ and $f_{\rm AGN}(\delta\sim0)$ in the summary table of \citet{2019ApJ...874...54M}, we calculate $R_{\rm AGN}$ for six types of the $z\sim2-3$ galaxy protoclusters which show the enhancement of AGN abundance. 

Figure \ref{fig_agn_merger_enhancement} presents the enhancement ratios of the galaxy merger fraction and the AGN fraction. We also show the fractional contribution of galaxy mergers relative to AGN, $R_{\rm merger}/R_{\rm AGN}$. The fractional contribution $R_{\rm merger}/R_{\rm AGN}$ varies among the galaxy protoclusters, ranging from $\sim10$\% to $\gtrsim100$\%. It is notable that the majority of galaxy protoclusters have $R_{\rm merger}/R_{\rm AGN}\gtrsim30$\%, except for BX/MD galaxies in QSO HS 1700$+$643 with $R_{\rm merger}/R_{\rm AGN}\sim10$\%. The high $R_{\rm merger}/R_{\rm AGN}$ fractions suggest that galaxy mergers play an important role in enhancing the AGN abundances in the galaxy protocluster at $z\sim2-3$. In contrast, in the galaxy protoclusters with low $R_{\rm merger}/R_{\rm AGN}$, other physical mechanisms might considerably contribute to the AGN enhancement, such as the smooth gas accretion to super massive black holes (e.g., \citealt{2005MNRAS.363....2K, 2009Natur.457..451D, 2011MNRAS.414.2458V}). The difference in $R_{\rm merger}/R_{\rm AGN}$ could also originate from cosmic variance due to the limited sample.

It should be noted that we simply compare the enhancement ratios of $R_{\rm merger}$ and $R_{\rm AGN}$ in this discussion. An agreement of $R_{\rm merger}$ and $R_{\rm AGN}$ does not necessarily imply a direct connection between galaxy mergers and the AGN enhancement in galaxy overdense regions. It is important to take into account several factors related to the AGN triggering processes, e.g., the time delay between galaxy mergers and the onset of gas accretion onto black holes (e.g., \citealt{2005MNRAS.361..776S, 2012MNRAS.420L...8H, 2023arXiv231208449C}). The purpose of this discussion is to demonstrate how the $R_{\rm merger}-\delta$ relation can be used to estimate quantitatively the contributions of physical mechanisms triggering the AGN activity. In addition, one might think that the super-resolution of AGN host galaxies provides more direct evidence of the physical connection between galaxy mergers and the AGN activity. However, we find it difficult to investigate the morphology of AGN host galaxies in the current super-resolution techniques, because faint structures of host galaxies are strongly affected by the light of central bright point sources. This is the reason why we have analyzed only the extended objects, i.e., galaxies, in this study (Section \ref{sec_selection}). Improvements of the super-resolution techniques for AGN host galaxies are left for future work. 

A similar analysis about the galaxy merger contribution can be conducted to various kinds of environmental dependences at high-$z$, for example, star formation rate (e.g., \citealt{2007A&A...468...33E,  2013MNRAS.434..423K, 2018MNRAS.473.1977S, 2019ApJ...887..214K, 2022A&A...662A..33L}), sub-millimeter galaxies (e.g., \citealt{2009Natur.459...61T, 2015ApJ...815L...8U}; see also \citealt{2018PASJ...70S..32U}), Ly$\alpha$ blobs (e.g., \citealt{2004AJ....128..569M, 2010ApJ...719.1654Y, 2008ApJ...678L..77P, 2017ApJ...837...71C}), and bright cluster galaxies (e.g., \citealt{2008ApJ...683L..17T, 2019ApJ...878...68I}). Our galaxy merger-overdensity relations would provide hints for unveiling the physical origins of these high-$z$ environmental dependences. 

% --------------------------------------------------
\section{Summary and Conclusions}\label{sec_summary}

We have super-resolved the seeing-limited Subaru HSC images in the three techniques: the classical RL PSF deconvolution (Section \ref{sec_classical}), the sparse modeling (Section \ref{sec_sparse}), and GAN (Section \ref{sec_gan}). By applying the super-resolution techniques to $32,187$ galaxies at $z\sim2-5$ selected in the survey areas of $\sim300$ deg$^2$ in HSC-SSP for $z\sim4-5$ and $\sim20$ deg$^2$ in CLAUDS for $z\sim2-3$, we have examined the environmental dependence of galaxy mergers at $z\sim2-5$. 

The major findings are summarized as follows: 

\vspace{0.5em}

\begin{enumerate}
  \item The three super-resolution techniques generate overall similar high spatial resolution HSC images (Figure \ref{fig_big_image}). In the qualitative and quantitative comparisons, we find that there are advantages and disadvantages in each super-resolution technique; for example, many noises tend to be seen in the classical RL PSF deconvolution, while the resolving powers of the sparse modeling and GAN are relatively low (Figures \ref{fig_other_imgs_hubble} and \ref{fig_tpr_fpr}). To alleviate the disadvantages of each super-resolution technique, we combine the three types of super-resolution images. The combined super-resolution images show galaxy substructures which closely resemble those seen in the Hubble images.
\vspace{0.5em}

  \item Using the combined super-resolution images, we measure the relative galaxy merger fraction corrected for the chance projection effect, $f_{\rm merger}^{\rm rel,\, col}$, at $z\sim2-5$ as a function of galaxy overdensity $\delta$. With the wide-area survey data and the super-resolution images, our $f_{\rm merger}^{\rm rel,\, col}$ measurements at $z\sim3$ validate the results of previous studies showing that $f_{\rm merger}^{\rm rel,\, col}$ is higher in higher-$\delta$ regions at $z\sim2-3$. In addition, we identify a trend that $f_{\rm merger}^{\rm rel,\, col}$ increases almost linearly with increasing $\delta$ at $z\sim4-5$  (Figure \ref{fig_delta_fmerg}). This is the highest-$z$ observational evidence that the galaxy merger fraction is related to $\delta$. The enhancement ratio of the galaxy merger fraction, $R_{\rm merger}$ (Equation \ref{eq_merger_enhancement}), shows a similar linear trend to the $f_{\rm merger}^{\rm rel,\, col}-\delta$ relation. In contrast to $z\gtrsim3$, we have not found a large $f_{\rm merger}^{\rm rel,\, col}$ increase with increasing $\delta$ at $z\sim2$, which might need wider survey data to confirm the $f_{\rm merger}^{\rm rel,\, col}-\delta$ relation at $z\sim2$.

\vspace{0.5em}

  \item We find that $f_{\rm merger}^{\rm rel,\, col}$ and $R_{\rm merger}$ measured in this study for $z\sim2-5$ and the literature for $z\sim2-3$ broadly align with the linear $f_{\rm merger}^{\rm rel,\, col}-\delta$ and $R_{\rm merger}-\delta$ relations (Figures \ref{fig_delta_fmerg_comparison} and \ref{fig_delta_fmerg_enhancement}). These alignments indicate that the environmental dependence of galaxy mergers does not largely change at least in a redshift range of $z\sim2-5$, allowing us to quantitatively estimate the contributions of galaxy mergers to various environmental dependences. In contrast, in massive galaxy clusters at $z\lesssim1$, the galaxy merger fraction is possibly suppressed compared to the $f_{\rm merger}^{\rm rel,\, col}-\delta$ and $R_{\rm merger}-\delta$ relations. This suppression is consistent with theoretical studies predicting that $f_{\rm merger}$ decreases in massive galaxy clusters at $z<1$ due to a high velocity dispersion of the low-$z$ massive galaxy clusters hindering galaxy-galaxy interactions. 

\vspace{0.5em}

  \item Using the $R_{\rm merger}-\delta$ relation, we estimate $R_{\rm merger}$ for $z\sim2-3$ galaxy protoclusters with high AGN abundances to investigate how galaxy mergers contribute to the enhancement ratio of the AGN fraction, $R_{\rm AGN}$ (Equation \ref{eq_agn_enhancement}). We find that the fraction of $R_{\rm merger}$ relative to $R_{\rm AGN}$ largely depends on the galaxy protoclusters, ranging from $R_{\rm merger}/R_{\rm AGN}\sim10$\% to $\gtrsim100$\%, but $R_{\rm merger}/R_{\rm AGN}\gtrsim30$\% for almost all of the galaxy protoclusters examined in this study (Figure \ref{fig_agn_merger_enhancement}). The high $R_{\rm merger}/R_{\rm AGN}$ fraction suggests that galaxy mergers play an important role in enhancing AGN abundances in galaxy protocluster regions. In contrast, in the galaxy protoclusters with a relatively low $R_{\rm merger}/R_{\rm AGN}$ fraction, other physical mechanisms may predominantly drive the AGN enhancement, such as the smooth cold gas accretion to galaxy overdense regions. 

\end{enumerate} 

\vspace{0.5em}

This super-resolution analysis has been made possible with the combination of the wide-area Subaru HSC and the high spatial resolution Hubble data. This method can be easily applied to the combinations of space telescopes, e.g., Euclid-JWST and Roman-JWST. These combinations would enable us to investigate morphological properties of rare galaxy populations at higher redshifts and on smaller scales than in the study with Subaru-Hubble. 

% ------
\begin{ack}
We thank the anonymous referee for valuable comments and suggestions that improved the manuscript. We acknowledge Rieko Momose, Kiyoaki Christopher Omori, Yoshiaki Ono, Taira Oogi, and Masayuki Yamaguchi for useful comments. This work was supported by JSPS KAKENHI Grant Numbers 20K14508, 22H01266, 22K14076, and 24H00002. 

The HSC collaboration includes the astronomical communities of Japan and Taiwan, and Princeton University. The HSC instrumentation and software were developed by the National Astronomical Observatory of Japan (NAOJ), the Kavli Institute for the Physics and Mathematics of the Universe (Kavli IPMU), the University of Tokyo, the High Energy Accelerator Research Organization (KEK), the Academia Sinica Institute for Astronomy and Astrophysics in Taiwan (ASIAA), and Princeton University. Funding was contributed by the FIRST program from Japanese Cabinet Office, the Ministry of Education, Culture, Sports, Science and Technology (MEXT), the Japan Society for the Promotion of Science (JSPS), Japan Science and Technology Agency (JST), the Toray Science Foundation, NAOJ, Kavli IPMU, KEK, ASIAA, and Princeton University. 

This paper makes use of software developed for the Large Synoptic Survey Telescope. We thank the LSST Project for making their code available as free software at  http://dm.lsst.org. 

This paper is based on data collected at the Subaru Telescope and retrieved from the HSC data archive system, which is operated by Subaru Telescope and Astronomy Data Center (ADC) at NAOJ. Data analysis was in part carried out with the cooperation of Center for Computational Astrophysics (CfCA), NAOJ.

The Pan-STARRS1 Surveys (PS1) have been made possible through contributions of the Institute for Astronomy, the University of Hawaii, the Pan-STARRS Project Office, the Max-Planck Society and its participating institutes, the Max Planck Institute for Astronomy, Heidelberg and the Max Planck Institute for Extraterrestrial Physics, Garching, The Johns Hopkins University, Durham University, the University of Edinburgh, Queen's University Belfast, the Harvard-Smithsonian Center for Astrophysics, the Las Cumbres Observatory Global Telescope Network Incorporated, the National Central University of Taiwan, the Space Telescope Science Institute, the National Aeronautics and Space Administration under Grant No. NNX08AR22G issued through the Planetary Science Division of the NASA Science Mission Directorate, the National Science Foundation under Grant No. AST-1238877, the University of Maryland, and Eotvos Lorand University (ELTE) and the Los Alamos National Laboratory.  

This research is based on data collected at Subaru Telescope, which is operated by the National Astronomical Observatory of Japan. We are honored and grateful for the opportunity of observing the  Universe from Maunakea, which has the cultural, historical and natural significance in Hawaii. 

These data were obtained and processed as part of the CFHT Large Area U-band Deep Survey (CLAUDS), which is a collaboration between astronomers from Canada, France, and China described in \citet{2019MNRAS.489.5202S}.  CLAUDS is based on observations obtained with MegaPrime/ MegaCam, a joint project of CFHT and CEA/DAPNIA, at the CFHT which is operated by the National Research Council (NRC) of Canada, the Institut National des Science de l???Univers of the Centre National de la Recherche Scientifique (CNRS) of France, and the University of Hawaii. CLAUDS uses data obtained in part through the Telescope Access Program (TAP), which has been funded by the National Astronomical Observatories, Chinese Academy of Sciences, and the Special Fund for Astronomy from the Ministry of Finance of China. CLAUDS uses data products from TERAPIX and the Canadian Astronomy Data Centre (CADC) and was carried out using resources from Compute Canada and Canadian Advanced Network For Astrophysical Research (CANFAR).

This research made use of SExtractor \citep{1996A&AS..117..393B}, SAOImage DS9 \citep{2003ASPC..295..489J}, Numpy \citep{2020Natur.585..357H}, Matplotlib \citep{2007CSE.....9...90H}, Scipy \citep{2020NatMe..17..261V}, Astropy \citep{2013A&A...558A..33A,2018AJ....156..123A},\footnote{http://www.astropy.org} and Ned Wright's Javascript Cosmology Calculator \citep{2006PASP..118.1711W}.\footnote{http://www.astro.ucla.edu/~wright/CosmoCalc.html}

\end{ack}

\bibliographystyle{aa}
\bibliography{bib_shibuya} 

% --------------------------------------------------
\section{Appendix}\label{sec_appendix}

\subsection{Other Example Galaxies}

\begin{figure}[th]
 \begin{center}
  \includegraphics[width=85mm]{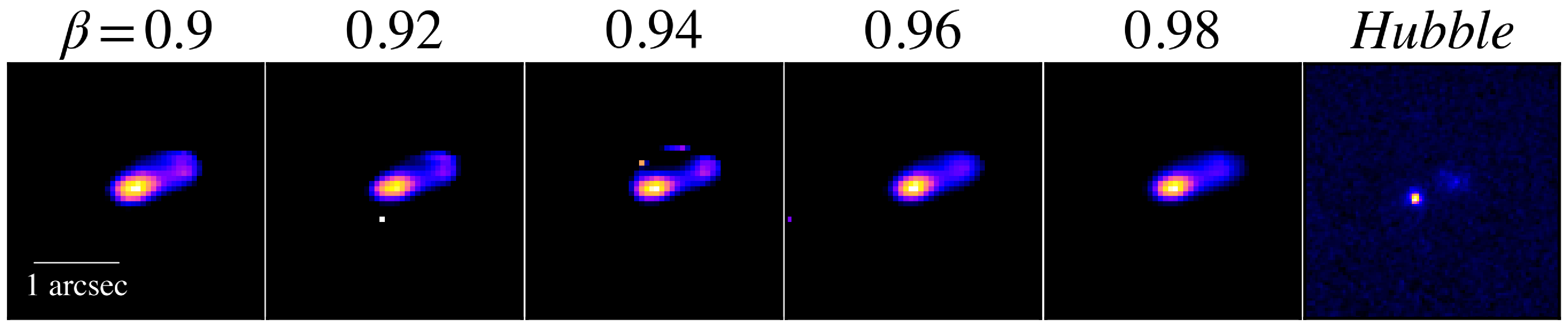}
  \includegraphics[width=85mm]{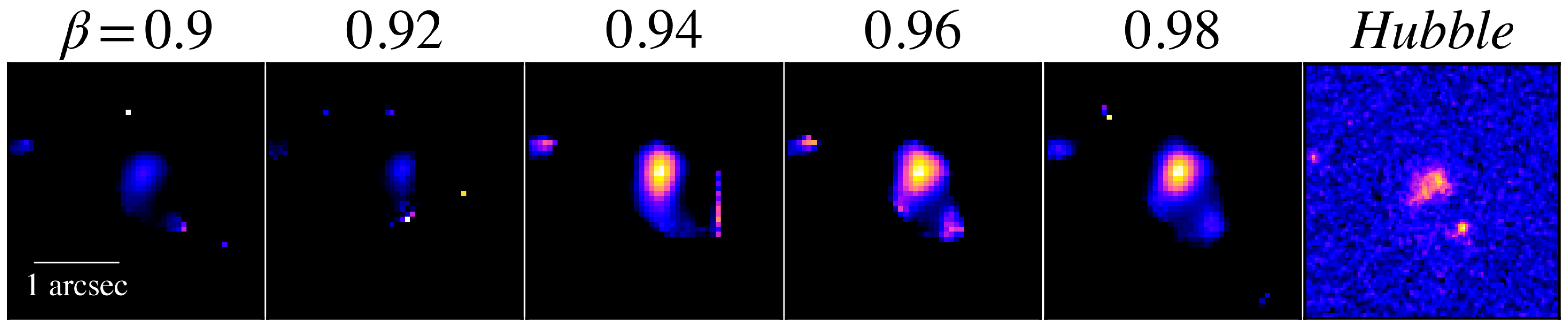}
  \includegraphics[width=85mm]{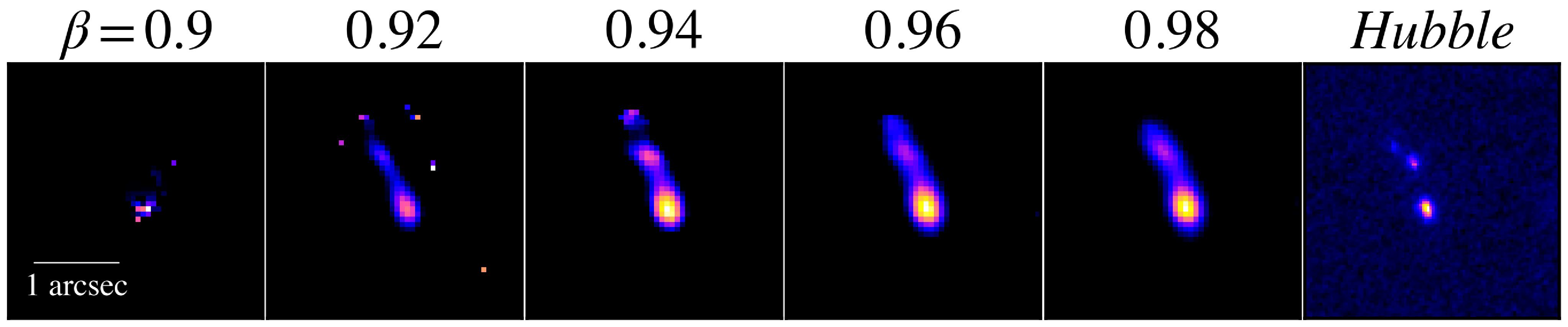}
 \end{center}
   \caption{Same as the top panel of Figure \ref{fig_params_sparse_gan}, but for other example galaxies.}\label{fig_params_sparse_others}
\end{figure}

\begin{figure}[th]
 \begin{center}
  \includegraphics[width=85mm]{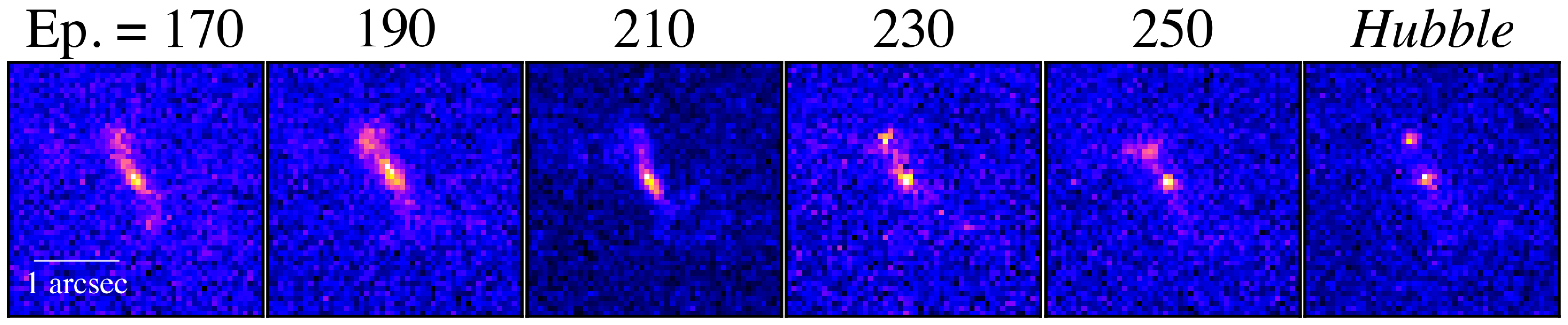}
  \includegraphics[width=85mm]{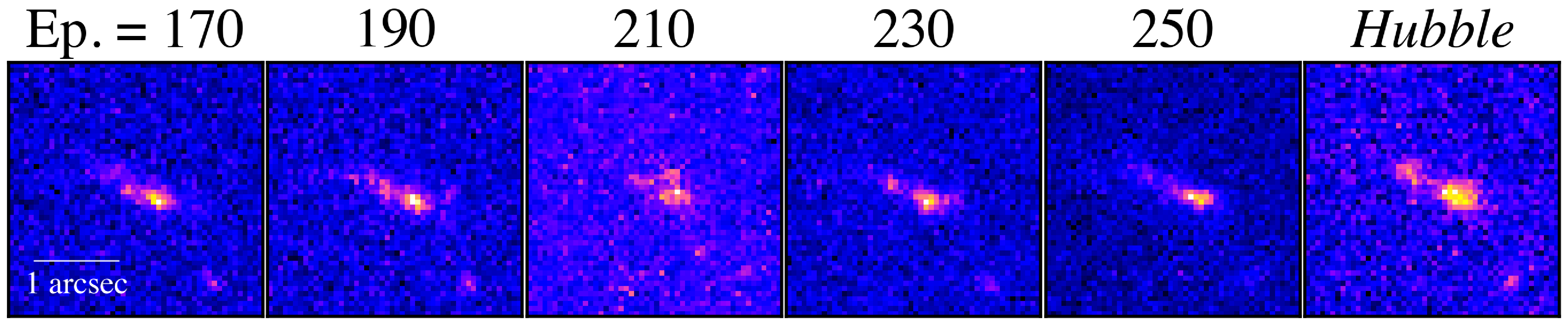}
  \includegraphics[width=85mm]{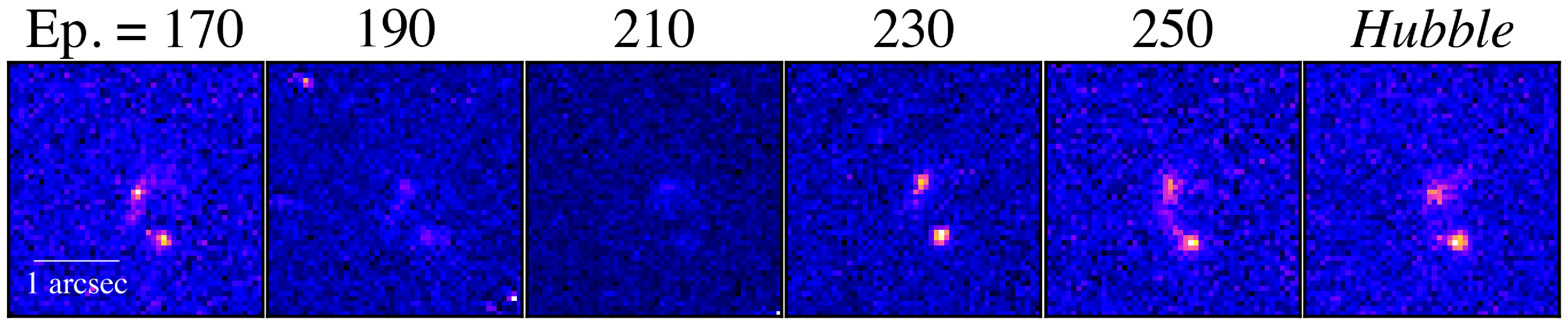}
 \end{center}
   \caption{Same as the bottom panel of Figure \ref{fig_params_sparse_gan}, but for other example galaxies.}\label{fig_params_gan_others}
\end{figure}

Figures \ref{fig_params_sparse_others} and \ref{fig_params_gan_others} show other example galaxies with different hyper parameters of sparse modeling and different epochs of GAN, respectively.

\subsection{Previous Observational Studies on the Relation between Galaxy Mergers and Galaxy Environments}

This section presents details of previous observational studies investigating the relation between galaxy mergers and galaxy environments. Table \ref{tab_observational} summarizes the redshift and the number of galaxies, the methods to identify galaxy mergers, and the results whether $f_{\rm merger}$ is related to $\delta$. 

\subsubsection{Hine et al. (2016)} 
  \citet{2016MNRAS.455.2363H} investigate the morphology of 23 Lyman-break galaxies (LBGs) in the SSA22 galaxy protocluster at $z\sim3.1$ and 86 coeval LBGs in the field environment. By visually classifying the LBGs using the Hubble ACS $I_{814}$-band images, which trace the rest-frame UV light at $\lambda\sim2,000$\,\AA, the study reveals a galaxy merger fraction of $0.48\pm0.10$, surpassing the $0.30\pm0.05$ observed in the field environment. This finding suggests an enhanced galaxy merger fraction in galaxy protocluster environments. For the plot in Figure \ref{fig_delta_fmerg}, we adopt $\delta=6$ as the galaxy overdensity of the SSA22 protocluster \citep{2000ApJ...532..170S} and $f_{\rm merger}$ of the `{\it combined field}' as the galaxy merger fraction in the field environment (Table 1 of \citealt{2016MNRAS.455.2363H}). 

\subsubsection{Monson et al. (2021)} 
  \citet{2021ApJ...919...51M} measure the galaxy merger fractions for eight LBGs and six AGN in the SSA22 galaxy protocluster and 49 field LBGs at $z\sim3.1$ using the Hubble WFC3 F160W images, which trace the rest-frame optical light at $\lambda\sim4,000$\,\AA. Morphological analyses with the S\'ersic profile fitting, the $G$-$M_{20}$ coefficients, and visual classifications suggest no significant differences in $f_{\rm merger}$ between the SSA22 galaxy protocluster and field LBGs. As for \citet{2016MNRAS.455.2363H}, we adopt $\delta=6$ as the galaxy overdensity of the SSA22 galaxy protocluster \citep{2000ApJ...532..170S} to plot $f_{\rm merger}$ in Figure \ref{fig_delta_fmerg}. We use $f_{\rm merger}$ of the `{\it Total Field}' and `{\it Protocluster}' as $f_{\rm merger}$ in the field and protocluster environments, respectively (Table 5 of \citealt{2021ApJ...919...51M}). 

\subsubsection{Mei et al. (2023)} 
  \citet{2023AA...670A..58M} visually classify galaxies in 16 spectroscopically confirmed galaxy clusters at $z\sim1.4-2.8$ from the Clusters Around Radio-Loud AGN (CARLA) survey \citep{2013ApJ...769...79W, 2014ApJ...786...17W} and field galaxies from the CANDELS survey. The total number of galaxies in the CARLA galaxy protoclusters is 271. The study finds a higher galaxy merger fraction of $f_{\rm merger}=0.26\pm0.03$ for the galaxy clusters than $f_{\rm merger}=0.17\pm0.05$ for the CANDELS fields, but no significant dependence of $f_{\rm merger}$ on the local environment within the CARLA sample. 

\subsubsection{Momose et al. (2022)} 
  \citet{2022arXiv221205984M} examine the relation between galaxy mergers at $z\sim2-2.5$ and the underlying 3D dark matter density reconstructed from galaxies and Ly$\alpha$ forest spectroscopic data. To measure $f_{\rm merger}$, the study cross-matches $1,344$ spectroscopically-confirmed galaxies with close-pair objects in galaxy merger catalogs \citep{2018ApJ...868...46S, 2020ApJ...904..107S}. The analysis suggests that $f_{\rm merger}$ does not clearly correlate with the 3D dark matter density. 
  
  \subsubsection{Liu et al. (2023)} 
  \citet{2023MNRAS.523.2422L} focus on two massive galaxy protoclusters, BOSS1244 and BOSS1542, at $z=2.24$. The galaxy overdensities are $\delta=5.6$ and $\delta=4.9$ for the BOSS1244 and BOSS1542 galaxy protoclusters, respectively. Using the Hubble WFC3 F160W data, the study investigates morphologies of 122 H$\alpha$ emitters in the galaxy protoclusters and 455 $z\sim2-3$ field galaxies from the CANDELS survey. Galaxy mergers are identified in the close-pair technique with $\mu\geq1/4$ and $d=5-30$ kpc. The galaxy merger fractions are $f_{\rm merger}=0.22\pm0.05$ in BOSS1244 and $f_{\rm merger}=0.33\pm0.06$ in BOSS1542, which are 1.8 and 2.8 times higher than those in the general fields, respectively. 

\subsubsection{Delahaye et al. (2017)} 
  \citet{2017ApJ...843..126D} compile photometric and spectroscopic samples of 55 galaxies in four galaxy clusters at $z=1.59-1.71$ from the SpARCS survey \citep{2009ApJ...698.1934M} and 91 field galaxies from the UKIDSS Deep Survey \citep{2007MNRAS.379.1599L}. The visual inspection is conducted with the Hubble NIR imaging data to identify galaxy mergers. The study finds that the galaxy merger fraction is $f_{\rm merger}=0.110^{+0.070}_{-0.056}$ for the galaxy cluster members, compared to $f_{\rm merger}=0.247^{+0.053}_{-0.046}$ for the field galaxies, ruling out the possibility of a mild or strong enhancement of the galaxy merger activity in galaxy clusters compared to the field.

\subsubsection{Coogan et al. (2018)} 
  \citet{2018MNRAS.479..703C} observe 11 galaxies in the galaxy cluster Cl J$1449+0856$ at $z=1.99$ using ALMA and JVLA. Visual inspections of the Hubble images, CO maps, and CO line profiles suggest that 5 out of the 11 galaxies are mergers or interacting systems, indicating a large number of mergers in the galaxy cluster core compared to the field environment. 

\subsubsection{Sazonova et al. (2020)} 
  \citet{2020ApJ...899...85S} study morphological properties for 498 galaxies in four galaxy clusters at $z=1.2-1.8$ using the Hubble WFC3 images. The study uses galaxies in CANDELS as a control sample. Based on morphological parameters of $A$ and $GM_{20}$ and close pair signatures, the researchers calculate the disturbance fractions and close neighbor fractions. The study finds that these fractions in the galaxy clusters are typically higher than those in the field. 

\subsubsection{Lotz et al. (2013)} 
  \citet{2013ApJ...773..154L} research a galaxy protocluster, IRC-0218A, at $z=1.62$ using the Hubble WFC3 F125W and F160W data. The galaxy overdensity of IRC-0218A is $\delta=20$. By identifying double nuclei or close pairs with $d<20$ kpc, the study measures $f_{\rm merger}$ for 14 galaxies in the galaxy protocluster and 130 galaxies at $z\sim1.6$ in the CANDELS fields. The $f_{\rm merger}$ measurements indicate that galaxy mergers in the galaxy protocluster are $\sim3-10$ times more abundant than in the field environment. For the plots in Figures \ref{fig_delta_fmerg_comparison} and \ref{fig_delta_fmerg_enhancement}, we use $f_{\rm merg}{\rm (cor)}$ of the `{\it Cluster all}' and `{\it Field $z\sim1.6$}' samples (Table 1 of \citealt{2013ApJ...773..154L}). 

\subsubsection{Watson et al. (2019)} 
  \citet{2019ApJ...874...63W} investigate two galaxy clusters, IRC0222A at $z\sim1.6$ and IRC0222B at $z\sim1.9$, identified through {\it Spitzer} mid-infrared observations. Using the Hubble F105W, F125W, and F160W band images, the study morphologically classifies 61 galaxies in the galaxy clusters and 888 field galaxies at $z\sim1.5-2.5$ in fields of COSMOS \citep{2007ApJS..172....1S} and CDFS \citep{2002ApJS..139..369G}. To identify galaxy mergers, the researchers combine visual inspections, the close pair technique, and selection criteria with grisim spectroscopic redshifts. The study finds galaxy merger fractions of $f_{\rm merger}=0.110^{+0.082}_{-0.032}$ in IRC0222A and $f_{\rm merger}=0.180^{+0.078}_{-0.045}$ in IRC0222B, which are higher than the measurement in the field, $f_{\rm merger}=0.050^{+0.011}_{-0.008}$. 
  
  \subsubsection{Laishram et al. (2024)} 
\citet{2024ApJ...964L..33L} examine the morphological features of [O II] emitters in extensive filamentary structures at $z\sim1.5$ discovered with a narrowband filter of Subaru HSC. The filamentary structures have a highly overdense core with a galaxy number density $\sim11$ times higher than the field average. By investigating the morphology of 368 galaxies with JWST NIRCam and the $GM_{20}$ indices, the study finds that galaxies in the dense core show a higher frequency of disturbances ($50\%\pm9\%$) compared to those in the outskirts ($41\%\pm9\%$) and the field ($21\%\pm5\%$), indicating more frequent mergers and interactions in denser environments.

\subsubsection{Lin et al. (2010)} 
  \citet{2010ApJ...718.1158L} explore galaxy mergers at $z=0.75-1.2$ using a sample of approximately $\sim20,000$ galaxies with reliable spectroscopic redshifts in the DEEP2 data \citep{2003SPIE.4834..161D}. The study identifies galaxy mergers in selection criteria of $d=10-50$ kpc/$h$ and a relative velocity difference, $|\Delta V|<500$ km/s. The galaxy overdensity is estimated in the 3rd-nearest neighbor method. The measurements of the galaxy merger fraction in low, med, and high $\delta$ bins suggest that the galaxy companion rate increases with galaxy overdensity, but after correcting for merger probability and timescales, wet galaxy mergers display weak environmental influence. In contrast, dry and mixed galaxy mergers are more effective in galaxy overdense regions. For the plots in Figures \ref{fig_delta_fmerg_comparison} and \ref{fig_delta_fmerg_enhancement}, we adopt $N_{\rm c}$ for the `{\it All}' type galaxy sample (Table 1 of \citet{2010ApJ...718.1158L}), which is most similar to our $f_{\rm merger}$ definition in the merger types. 
   
\subsubsection{Kampczyk et al. (2013)} 
  \citet{2013ApJ...762...43K} examine the relation between galaxy mergers and galaxy environments at $z=0.2-1$ using the zCOSMOS 10k-bright spectroscopic sample \citep{2007ApJS..172...70L, 2009ApJS..184..218L}. The study identifies 153 close kinematic pairs out of 3,667 galaxies in selection criteria of $d<100$, $50$, or $30$ kpc/$h$, $|\Delta V|<500$ km/s, and $\mu\geq1/4$. The galaxy overdensity is estimated by counting the number of galaxies within radius-fixed apertures and source velocity differences $|\Delta V|<1,000$ km/s. By correcting for projection effects, the study reveals a three times higher pair fraction in the top density quartile compared to the lowest, indicating a higher galaxy merger rate in denser galaxy environments. 

\subsubsection{de Ravel et al. (2011)} 
  \citet{2011arXiv1104.5470D} analyze 10,644 galaxies at $z=0.2-1$ from the zCOSMOS spectroscopic survey. The research identifies 263 close pairs with selection criteria of $d<100$, $50$, $30$, or $20$ kpc/$h$, $|\Delta V|<500$ km/s, and $\mu\geq1/4$. The galaxy overdensity is estimated by counting the number of galaxies within radius-fixed apertures. The pair fraction and merger rate show an increase with local galaxy density, suggesting that dense galaxy environments favor major merger events. Although there is no $f_{\rm merger}-\delta$ data available in the paper, based on the description of `{\it ... for an environment 10 times denser, one finds 3.0, 2.6 and 2.1 times more pairs for separations of $r = 100, 50$ and $30$ h$^{-1}$ kpc respectively.}', we plot $R_{\rm merger}$ in Figure \ref{fig_delta_fmerg_enhancement}. 

\subsubsection{Kocevski et al. (2011)} 
  \citet{2011ApJ...736...38K} investigate morphologies of galaxies in three clusters and three groups of the Cl1604 galaxy supercluster at $z\sim0.9$ using the Hubble ACS images. The study classifies 131 galaxies in the core regions within the virial radii of the six overdense systems, $R<R_{\rm vir}$, and 70 galaxies at the outer regions, $R>R_{\rm vir}$. By combining visual inspections with results of the $GM_{20}$ classification, the study finds that the fractions of galaxies with disturbed morphologies are typically higher at $R<R_{\rm vir}$ than those at $R>R_{\rm vir}$, indicating that galaxy mergers are more prevalent in higher galaxy density regions. 

\subsubsection{Asano et al. (2020)}
  \citet{2020ApJ...899...64A} measure the galaxy merger fractions in four environments within and around the galaxy cluster CL1604-D at $z=0.923$: the field, SW group, NE group, and cluster core. By selecting mergers/interacting galaxies in the same way as in \citet{2011ApJ...736...38K}, the study finds that the galaxy merger fraction varies across the regions. The galaxy mergers are scarce in the field and the cluster core ($f_{\rm merger}\sim8$\%). In contrast, merger/interacting galaxies are more prevalent in the galaxy group environments, the SW and NE groups ($f_{\rm merger}\sim20$\% and $50$\%, respectively). 

\subsubsection{Paulino-Afonso et al. (2019)}  
  \citet{2019AA...630A..57P} examine galaxy structures for 490 galaxies in and around a galaxy superstructure at $z\sim0.84$ in the COSMOS field. With the Hubble ACS $I_{814}$-band images in COSMOS, the study measures the morphological parameters of $CAS$ and $GM_{20}$ and uses visual classification catalogs of \citet{2017MNRAS.464.4176W} to investigate environment impacts of galaxy structures. The galaxy overdensity is estimated based on the 3D position of the spectroscopically confirmed galaxies. The study finds that the Asymmetry parameter $A$ and the fraction of irregular galaxies are nearly constant with the galaxy overdensity. However, increasing and decreasing trends are found if the sample is divided into star-forming galaxies (SFGs) and quiescent galaxies (QGs). The fraction of QGs (SFGs) with irregular morphologies increases (decreases) from the field/filament-like regions to the highest galaxy density environment. 

\subsubsection{van Dokkum et al. (1999)} 
  \citet{1999ApJ...520L..95V} conduct a morphological analysis of galaxies in a luminous X-ray cluster, MS 1054$-$03 at $z = 0.83$. Examining $V_{606}$ and $I_{814}$ images of the Hubble WFPC2 for 81 spectroscopically confirmed cluster members, the researchers identify 13 ongoing mergers. This means that the galaxy merger fraction is $f_{\rm merger}=17$\% for the $L\gtrsim L^*$ cluster population, which is higher than those for the field environments and lower-$z$ rich galaxy clusters. These galaxy mergers are preferentially found in the outskirts of the MS 1054$-$03 galaxy cluster. In addition to MS 1054$-$03, they measure $f_{\rm merger}$ for a lower-$z$ galaxy cluster, CL 1358$+$62 at $z=0.33$ \citep{1998ApJ...500..714V}. The galaxy merger fraction in CL 1358$+$62 is $f_{\rm merger}\sim0.005$, much lower than the $f_{\rm merger}$ value in MS 1054$-$03. 

\subsubsection{Omori et al. (2023)} 
  \citet{2023AA...679A.142O} classify 302,148 galaxies at $z\sim0.01-0.35$ in the Subaru HSC-SSP survey using the Zoobot deep learning model pre-trained on the Galaxy Zoo DeCALS data. The Zoobot model is fine-tuned with synthetic HSC images from the TNG simulation. The study identifies that at scales of $0.5-8$ $h^{-1}$ Mpc, merger galaxies prefer lower density environments, while non-mergers favor higher density environments. However, below these scales, the TNG simulation data shows a reversal in the trend, with higher density environments favoring galaxy mergers. 

\subsubsection{Sureshkumar et al. (2024)}
\citet{2024arXiv240218520S} determine whether galaxy mergers at $z=0.1-0.15$ occur more frequently in dense or less dense environments. Using data from the GAMA survey, the study classifies 23,855 galaxies into mergers and non-mergers based on the convolutional neural network and morphological parameters of $GM_{20}$. While no significant difference is found in the clustering strengths of mergers and non-mergers using the two-point correlation function, an anti-correlation is observed between the likelihood of a galaxy being a merger and the galaxy environment using the marked correlation function. The results suggest that galaxy mergers prefer under-dense galaxy environments. 

\subsubsection{McIntosh et al. (2008)} 
  \citet{2008MNRAS.388.1537M} analyze 5,376 member galaxies in galaxy groups and clusters at $z\leq0.12$. By identifying close pairs with $\leq30$ kpc projected separations and asymmetric structures in the S\'ersic-fit residuals of the SDSS images, the study isolates 38 mergers. Investigating the galaxy merger frequency as a function of dark matter halo mass, the researchers find that galaxy mergers are more likely to occur in large galaxy groups than in galaxy clusters. 

\subsubsection{Liu et al. (2009)} 
  \citet{2009MNRAS.396.2003L} identify 18 major dry mergers out of 515 early-type Brightest Cluster Galaxies (BCGs) at $z=0.03-0.12$. The galaxy major mergers are selected as galaxy pairs with projected separations of $<30$ kpc and with signatures of interaction in SDSS residual images. The study finds that the galaxy major merger fraction in the BCGs increases with cluster richness. 

\subsubsection{Alonso et al. (2012)}
\citet{2012AA...539A..46A} examine 660 galaxy pairs in galaxy groups at $z<0.1$ from SDSS-DR7. The galaxy pairs are categorized into `Merging', `Tidal', and `Non-disturbed' pairs based on levels of galaxy interactions. The galaxy pairs are concentrated towards group centers, with the disturbed pairs often containing the brightest galaxies in the galaxy groups. To analyze the influence of the group global environment, the density of the galaxy pairs are quantified using the distance to the fifth nearest neighbor. Galaxy pairs have an excess of the low-density group global environments compared to the control sample. This trend is more pronounced in the merging pairs than in the tidal pairs and non-disturbed pairs. 

\subsubsection{Perez et al. (2009)}  
  \citet{2009MNRAS.399.1157P} use galaxies at $z=0.01-0.1$ in the SDSS DR4 catalog to investigate the relation between galaxy mergers and galaxy environments. The study defines close pairs with a projected separation of $<100$ kpc $h^{-1}$ and a relative velocity difference of $\Delta V<350$ km/s as galaxy pairs. Based on visual inspections for the sample of galaxy pairs, the researchers also identify merging systems with morphological disturbances and strong signals of interactions. The galaxy overdensity is calculated by using the projected distance to the fifth nearest neighbor of galaxies. The study analyzes the cumulative fractions of galaxy mergers and isolated galaxies in different galaxy overdensities. The cumulative fractions of galaxies suggest that the numbers of close galaxy pairs and merging systems increase in intermediate-density environments. 

\subsubsection{Ellison et al. (2010)}
\citet{2010MNRAS.407.1514E} examine galaxy mergers at $z<0.1$ in various galaxy environments using 5,784 close galaxy pairs from the SDSS DR4 data. The analysis of the galaxy merger fraction indicates that galaxy mergers occur across all the galaxy densities. The low galaxy density areas favor triggered star formation, while high galaxy density regions show galaxy mergers without star formation. Measurements of the Asymmetry index reveal a high galaxy merger fraction in cluster centers. 

\subsubsection{Darg et al. (2010)}
\citet{2010MNRAS.401.1552D} compare the galaxy local densities of galaxy pairs and control galaxies at $z=0.005-0.1$ from SDSS DR6 to understand galaxy merger properties and galaxy  environmental effects. The galaxy pairs are identified using the Galaxy Zoo classification catalog. The study finds no significant difference in the environmental distribution between the galaxy mergers and the control galaxies. Elliptical galaxies in galaxy mergers tend to occupy slightly denser galaxy environments than the controls, while spiral galaxies show a minimal difference. 

\subsubsection{Moss (2006)} 
  \citet{2006MNRAS.373..167M} measures the galaxy merger fraction for 680 galaxies in eight galaxy clusters at $z\sim0$. The galaxies are likely to reside in intermediate-density infall regions according to the velocity distribution of the galaxies. Visual inspections are conducted to classify the galaxies based on the presence of disturbance signatures. The analysis shows that $\sim50-70$\% of the galaxies in the infall population are interacting or merging systems. The galaxy merger fraction is $\sim4-5$ times higher than $\sim13$\% for the field galaxies located outside the virial radius of the galaxy clusters. 
  
\end{document}